\documentstyle[epsbox,epsfig,12pt]{article}
\newcommand{\bold}[1]{\mbox{\boldmath ${#1}$}}

\begin{document}
\baselineskip 4 ex
\title{The relation between the photonuclear E1 sum rule and the effective orbital g-factor 
\thanks{Correspondence to: W. Bentz, 
e-mail: bentz@keyaki.cc.u-tokai.ac.jp}}
\author{Wolfgang Bentz \\
Dept. of Physics, School of Science, \\
Tokai University \\
1117 Kita-Kaname, Hiratsuka 259-1207, Japan \\
{ }\\
and \\
Akito Arima \\
House of Councillors \\
2-1-1 Nagata-cho, Tokyo 100-8962, Japan}

\date{}
\maketitle
\begin{abstract}
The connection between the enhancement factor ($1+\kappa$) of
the photonuclear E1 sum rule and the orbital angular momentum g-factor ($g_{\ell}$) of
a bound nucleon is investigated in the framework of the Landau-Migdal
theory for isospin asymmetric nuclear matter. Special emphasis is put on the role of gauge invariance
to establish the $\kappa-g_{\ell}$ relation. By identifying the physical processes
which are taken into account in $\kappa$ and $g_{\ell}$, the validity and limitations of this
relation is discussed. The connections to the collective excitations and to nuclear Compton 
scattering are also shown.   
\end{abstract}

\newpage

\section{Introduction}
The enhancement factor ($1+\kappa$) of the photonuclear E1 sum rule and the orbital angular 
momentum g-factor ($g_{\ell}$) of a bound nucleon 
have attracted the attention of nuclear theorists as well as experimentalists for
a long time, since these quantities reflect the presence of exchange forces
and mesonic degrees of freedom in nuclei 
\cite{LEB,MIY,FUI,YAM}. More than 30 years ago, Fujita and Hirata \cite{FUH} used the 
isospin symmetric Fermi
gas as a model for an N=Z nucleus to derive the simple relation 
$1+\kappa=2\,g_{\ell,{\rm IV}}$ between $\kappa$ and the isovector (IV) part 
of $g_{\ell}$ in first order perturbation theory. Later it has been shown
\cite{ABHI} that, because of the presence of correlations between the nucleons,
only a part of the total $\kappa$ is related to $g_{\ell,{\rm IV}}$. It has been argued \cite{FIH} that
this part of $\kappa$ is related to the sum of the E1 strength in the region of the isovector 
giant dipole resonance (GDR). In more recent years \cite{SCHU2}, this modified $\kappa-g_{\ell}$ relation 
has been used to analyse the results of photo-neutron experiments
\cite{PNE} and photon scattering experiments \cite{SCHU1,DAL}, in particular for nuclei in
the lead region. In these analyses, the corrections to $g_{\ell,{\rm IV}}$ were associated
with the meson exchange currents. 

On the other hand, as early as 1965, Migdal and collaborators \cite{MIG} used an approach
based on a a gas of quasiparticles to relate $\kappa$ to the parameters 
characterizing the interaction between the quasiparticles 
(the Landau-Migdal parameters). Combining this relation with the more general one 
between $g_{\ell,{\rm IV}}$ and the Landau-Migdal parameters \cite{MIGB},  
their approach suggested that the relation $1+\kappa=2\,g_{\ell,{\rm IV}}$ holds more
generally without recourse to perturbation theory.   
The fact that their result involves the total $\kappa$ instead of just a part of it
reflects the quasiparticle gas approximation. 
Concerning the orbital g-factor, however, the Landau-Migdal approach suggests that it is the 
{\em total} $g_{\ell,{\rm IV}}$, and not only the part associated with the meson exchange currents, 
which enters in the relation to the E1 enhancement factor. By using a somewhat different approach,
this result has also been obtained by Ichimura \cite{ICHI}. 

The main advantage of the Landau-Migdal theory \cite{MIGB}, which is based on the Fermi liquid
approach due to Landau \cite{LAND,NOZ}, is
that symmetries, like gauge invariance and Galilei invariance, are incorporated rigorously,
and that the description of the collective excitations of the system is physically very
appealing. The Fermi liquid approach to discuss sum rules in nuclear matter has therefore 
turned out to be very fruitful, and has been used in several papers on giant resonances
\cite{LS,SW}. However, to the best of our knowledge, a general
discussion of the physical processes which contribute to the $\kappa$-$g_{\ell}$ relation,
as well as a discussion on the physical origin of this relation, is still outstanding.  

The strong interest in nuclear giant resonances and sum rules is continuing nowadays \cite{CONF,BOOK},
and new phenomena like the double giant resonances \cite{DGR} or dipole resonances in
neutron rich nuclei \cite{NRN} have attracted attention. The Landau-Migdal theory, which is a strong
candidate to describe the giant resonances \cite{KST}, is now extended in various directions \cite{SPETH} 
so as to give a more general description which is valid up to higher excitation
energies. In the light of these recent developments and the analyses of photonuclear
experiments mentioned above, more detailed understanding on the $\kappa$-$g_{\ell}$ relation
would be desirable.   

In this paper we will present a general discussion on the
$\kappa$-$g_{\ell}$ relation in isospin asymmetric nuclear matter, using the language of the 
Landau-Migdal theory. The aims of our work are as follows: First, we will extend the
relations obtained previously for the orbital g-factor and the E1 enhancement factor to the case 
of $N \neq Z$, putting special emphasis on the role of gauge invariance. Second, we will identify 
the physical processes which are taken into account in the $\kappa$-$g_{\ell}$ relation,
both in terms of Feynman diagrams as well as time-ordered Bethe-Goldstone diagrams.
In this connection we will also discuss the relation to the collective excitations. 
Third, we will show the connection to the photon scattering amplitude, which establishes the
physical origin of the $\kappa$-$g_{\ell}$ relation.       

The rest of the paper is organized as follows: In sect. 2 we briefly review  some
relations of the Landau-Migdal theory which will be used for the discussions on
$g_{\ell}$ and $\kappa$. In sect. 3 we derive the expressions for the proton and
neutron orbital g-factors in isospin asymmetric nuclear matter, which are in principle exact 
and hold also in relativistic
field theory. In sect. 4 we discuss the E1 sum rule and its relation to the orbital g-factors, limiting 
ourselves to the nonrelativistic case because of the problems associated with the center of mass
motion. In sect. 5 we discuss the physical processes which are taken into account in the
$\kappa-g_{\ell}$ relation and establish the connection to the collective excitations.
In sect. 6 we discuss the relation to the nuclear Compton scattering amplitude, 
which provides the physical origin of the $\kappa$-$g_{\ell}$ relation. A summary and conclusions
are given in sect. 7.  

\section{Vertices and correlation functions in the Landau-Migdal theory}
\setcounter{equation}{0}
In this section we briefly review some relations of the Landau-Migdal theory 
which will be used in later sections.

\begin{figure}[h]
\begin{center}
\epsfig{file=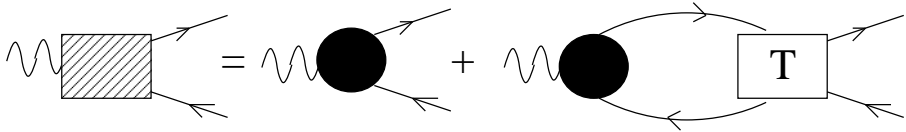,angle=0,width=12cm}
\caption{Graphical representation of relation (\ref{gam}) for the vertex.
The shaded square stands for the full vertex ($\Gamma$), and the black circle for
the vertex which is two-body irreducible in the p-h channel ($\Gamma_0$). $T$ denotes the $T$-matrix.}
\end{center}
\end{figure}

The integral equation for the electromagnetic vertex $\Gamma$ is graphically shown in
Fig. 1 and is written symbolically \footnote{In the symbolic notations we will omit the
Lorentz indices for the electromagnetic vertices, currents and correlation functions.} as
\begin{equation}
\Gamma=\Gamma_0 -i\,T\,S\,S\,\Gamma_0,
\label{gam}
\end{equation}
where $\Gamma_0$ is the part which is two-body irreducible in the particle-hole (p-h) channel, $S$ is the full nucleon
propagator, and $T$ is the two-body T-matrix which satisfies the Bethe-Salpeter equation
\begin{equation}
T=K -i\, T\,S\,S\,K
\label{bs}
\end{equation}
with the two-body irreducible kernel $K$. The important step in the renormalization procedure of the
Landau-Migdal theory is to split the product $S\,S$ into the two parts \cite{MIGB}
\begin{equation}
S\,S\,=A\,+\,B,
\label{split}
\end{equation}
where $A$ denotes the product of the pole parts of the particle and hole propagators
(${\displaystyle A=S_p^{\rm (pole)} S_h^{\rm (pole)}}$), including
the prescription to evaluate the other energy dependent quantities in a frequency loop integral  
for on-shell p-h states. (The explicit form of $A$ in the momentum representation
will be specified below.) All other parts, like non-pole parts, the product of two particle or two hole
propagators, antinucleon propagators etc, are included in the quantity $B$. The basic idea here is that
the part $A$ represents the ``active space'', while the effects of $B$ are renormalized into the effective
vertex and effective interaction. 

The equation (\ref{bs}) for the T-matrix is then equivalent to
\begin{eqnarray}
T&=&T^{(\omega)}-i\, T\,A\,T^{(\omega)} \label{tfu} \\
T^{(\omega)}&=&K -i\, T^{(\omega)}\,B\,K \label{tom},
\end{eqnarray}
and equation (\ref{gam}) for the vertex is equivalent to
\begin{eqnarray}
\Gamma&=&\Gamma^{(\omega)} - i T^{(\omega)}\,A\,\Gamma  \label{gfu} \\
\Gamma^{(\omega)}&=&\Gamma_0 -i T^{(\omega)}\,B\,\Gamma_0.
\label{gom}
\end{eqnarray} 
As is clear from these equations, the quantities $T^{(\omega)}$ and $\Gamma^{(\omega)}$ 
do not involve the product $S_p^{\rm pole} S_h^{\rm pole}$ in the intermediate states, that is, they
have no p-h cuts, and eqs. (\ref{tfu}) 
and (\ref{gfu}) can be considered
as RPA-type equations with $T^{(\omega)}$ and $\Gamma^{(\omega)}$ playing the role of
the effective interaction and the effective vertex, respectively, acting in the space of p-h states. 

\begin{figure}[h]
\begin{center}
\epsfig{file=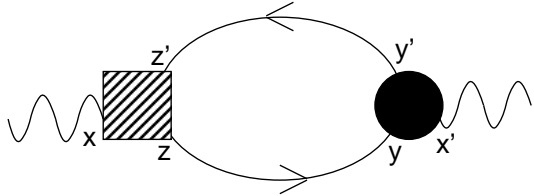,angle=0,width=7cm}
\caption{Graphical representation of eq.(\ref{pol}). For explanation of the symbols, see the
caption to Fig.1.}
\end{center}
\end{figure}

Another quantity which is of particular interest for the collective excitations of the
system is the electromagnetic current-current correlation function, or shortly the
correlation function, which can be expressed in coordinate space as follows (see Fig.2)
\footnote{We note
that, although this diagram is drawn in the p-h channel, no particular time 
ordering has been chosen, e.g; the product $S\,S$ in (\ref{pol})
can also involve two particle (forward propagating) Green functions.}:
\begin{equation}
i\,\langle 0|T\left(j^{\mu}(x')j^{\nu}(x)\right)|0\rangle =
i\,{\rm Tr}\,\, \left[\Gamma_0^{\mu}(y,x',y')\,S(y',z')\,\Gamma^{\nu}(z',x,z) \,S(z,y)\,\right],
\label{pol}
\end{equation} 
or symbolically in the form
\begin{equation}
\Pi = i\,\Gamma_0\,S\,S\,\Gamma  \label{polsym}
\end{equation}
The trace (${\rm Tr}$) in (\ref{pol}) stands for an integral over the space-time positions $z',\,z,\,
y',\,y$ as well as the trace over spin-isospin indices.  

If we use eqs.
(\ref{split}),$\,$ (\ref{gfu}) and (\ref{gom}) in eq.(\ref{polsym}), we obtain the separation
of the correlation function into a part $\Pi_A$, which involves p-h cuts, and a part
$\Pi_B$, which involves the rest, like $2p-2h$ cuts etc.:
\begin{equation}
\Pi = i \,\Gamma^{(\omega)}\,A\, \Gamma \,+\,i \,\Gamma_0\, B\, \Gamma^{(\omega)} 
\equiv \Pi_A + \Pi_B. \label{res}
\end{equation}
The use of eq.(\ref{gfu}) for the vertex $\Gamma$ then generates the RPA series for the
part $\Pi_A$. 

For illustration, we represent $\Pi_A$ by Fig. 3, and show some selected examples
\footnote{The diagrams of Figs. 4 and 5 should be understood as examples for time-ordered diagrams 
in the context of a mean field approximation, like the Hartree-Fock approximation, for the single 
particle Green functions. 
As we will discuss in detail in sect. 5, the 2p-2h diagrams shown here give contributions to
{\em both} $\Pi_A$ and $\Pi_B$. Examples for self energy corrections are not shown in this figure, but
are also discussed in sect.5.} 
for diagrams which contribute to $\Pi_A$ in Fig. 4.

\begin{figure}[h]
\begin{center}
\epsfig{file=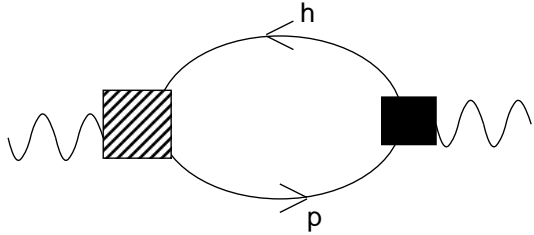,angle=0,width=7cm}
\caption{Graphical representation of the part $\Pi_A$ in eq.(\ref{res}). The shaded
square represents $\Gamma$, the black square $\Gamma^{(\omega)}$, and p (h) denotes 
the pole part of the particle (hole) propagator.}
\end{center}
\end{figure}

\begin{figure}[h]
\begin{center}
\epsfig{file=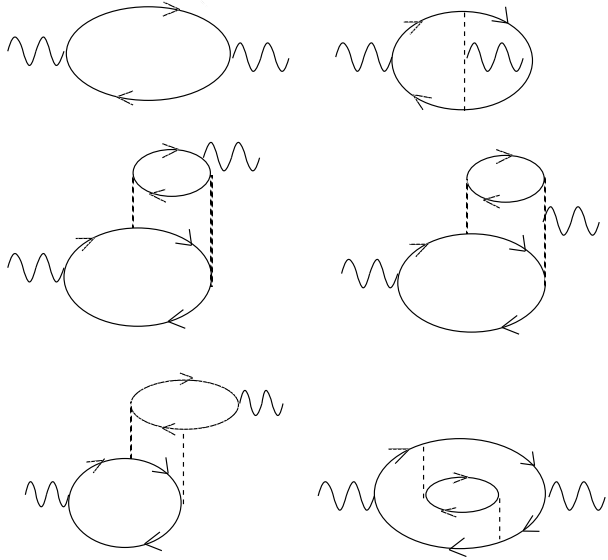,angle=0,width=8cm}
\caption{Some selected time-ordered diagrams contributing to $\Pi_A$. The first line shows the 
non-interacting part and a meson-exchange current contribution in first order perturbation 
theory, the second line shows some graphs which arise from 2p-2h states in the vertex
$\Gamma^{(\omega)}$, and the third line shows some examples reflecting the 2p-2h states
in the interaction $T^{(\omega)}$. The vertex where the 
external field couples to a dashed line stands symbolically for the two-body exchange currents. 
Self energy corrections also contribute, but are not shown in this figure.}
\end{center}
\end{figure}

The common feature of these diagrams
is that they involve at least one electromagnetic vertex where a p-h pair is created by the
external field. This is in contrast to the diagrams contributing exclusively to  
$\Pi_B$, for which we show some examples in Fig. 5.  

\begin{figure}[h]
\begin{center}
\epsfig{file=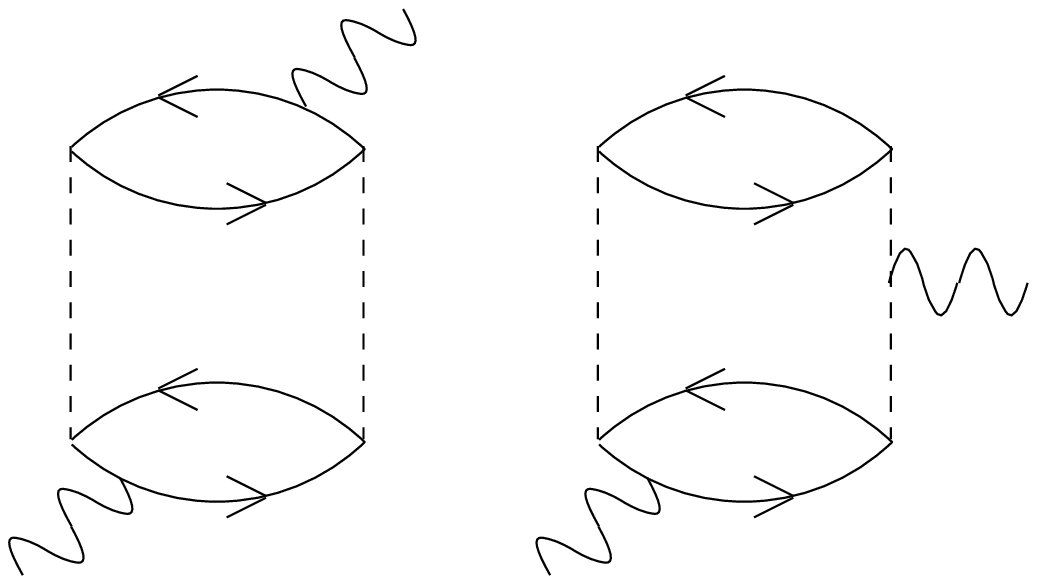,angle=0,width=8cm}
\caption{Some selected time-ordered diagrams which contribute exclusively to $\Pi_B$.}
\end{center}
\end{figure}

We now specify the form of $A$ in momentum space, which corresponds to the product of the pole parts
of particle and hole propagators:
\begin{eqnarray}
S_p^{\rm (pole)}(k) = \frac{Z_k\,(1-n_k)}{k_0-\epsilon_k +i\delta}\,,\,\,\,\,\,\,\,\,\,\,\,\,\,\,\,\,\,\,\,   
S_h^{\rm (pole)}(k) = \frac{Z_k\,n_k}{k_0-\epsilon_k -i\delta}\,, \label{sk} 
\end{eqnarray}
where $\epsilon_k$ and $Z_k$ are the quasiparticle poles and residues, and $n_k$ is the
Fermi distribution function. From the structure of the RPA-type equations (\ref{tfu}) and
(\ref{gfu}) it is clear that each loop integral is associated with one factor
$A$. We therefore consider the following expression to represent a generic loop integral
which involves a p-h cut:
\begin{eqnarray}
{\cal I} &=& \int \frac{{\rm d}^4 k}{(2\pi)^4} \left(S_p^{(\rm pole)}(k+\frac{q}{2})\,S_h^{(\rm pole)}
(k-\frac{q}{2}) + S_h^{(\rm pole)}(k+\frac{q}{2})\,S_p^{(\rm pole)}(k-\frac{q}{2}) \right) \nonumber \\
& \times & F(k+\frac{q}{2},k-\frac{q}{2}) \,,
\label{loop}
\end{eqnarray}
where $F$ does not contain $A$, and
$q^{\mu}=(\omega,{\bold q})$ is the momentum transferred by the external field. 
By closing the $k_0$ contour in the upper plane, we get
\begin{eqnarray}
{\cal I} &=& i \int \frac{{\rm d}^3 k}{(2\pi)^3} \,Z_{k+} Z_{k-}\,
\left(\frac{(1-n_{k+})n_{k-}}{\omega - (\epsilon_{k+} - \epsilon_{k-}) + i \gamma} 
F\left((\omega+\epsilon_{k-})\,{\bold k}_+ \, ; \, \epsilon_{k-}\,{\bold k}_-\right) \right. \nonumber \\  
&+& \left.\frac{n_{k+} (1-n_{k-})}{\omega - (\epsilon_{k+} - \epsilon_{k-}) + i \gamma}
F\left(\epsilon_{k+}\,{\bold k}_+\, ; \, (-\omega+\epsilon_{k+})\,{\bold k}_-\right) \right) \label{loop1}
+ {\cal I}_F\,\,,
\end{eqnarray}
where $\gamma\equiv\,$sign$\,\omega$, ${\cal I}_F$ denotes the contributions from the poles of $F$ in the upper plane, and
all quantities with subindex ${k\pm}$ refer to the three momentum
${\bold k}_{\pm}={\bold k}\pm{\bold q}/2$. We expand both
functions $F$ in (\ref{loop1}) around $\omega=\epsilon_{k+}-\epsilon_{k-}$, and take only the
first term in this expansion. This, together with the prescription to omit the term ${\cal I}_F$, {\em defines}
the quantity $A(k,q)$:
\begin{eqnarray}
{\cal I}_A &=& -i \int \frac{{\rm d}^3 k}{(2\pi)^3}\, \frac{(n_{k+}-n_{k-}) \,Z_{k+} Z_{k-}}
{\omega - (\epsilon_{k+} - \epsilon_{k-}) + i \gamma}\, F(\epsilon_{k+}\,{\bold k}_+\, ; \, \epsilon_{k-}\, 
{\bold k}_-)
\label{loop2} \\
&\equiv& \int \frac{{\rm d}^4 k}{(2\pi)^4} \, A(k,q)\,F(k+\frac{q}{2},k-\frac{q}{2}). \label{def} 
\end{eqnarray}
From this identification we obtain 
\begin{eqnarray}
A(k,q)  = - 2\pi i \delta\left(k_0 - \frac{\epsilon_{k+} + \epsilon_{k-}}{2}\right) 
\frac{(n_{k+}-n_{k-}) \,Z_{k+} Z_{k-}}{\omega - (\epsilon_{k+} - \epsilon_{k-}) + i \gamma}
\, {\hat P}(\omega)\,, \label{a}
\end{eqnarray}
where ${\hat P}(\omega)$ denotes the prescription to evaluate $F$ at $\omega=\epsilon_{k+}-\epsilon_{k-}$.
Together with the delta function in (\ref{a}) this means that $A$ effectively 
replaces the function ${\displaystyle F(k+\frac{q}{2},k-\frac{q}{2})}$ by its ``on-shell'' (o.s.) value defined by
\begin{eqnarray}
k_0 +\frac{\omega}{2} \stackrel{{\rm o.s.}}{=} \epsilon_{k+},\,\,\,\,\,\,\,\,\,\,\,\,\,\,\,\,\,\,\,\, 
k_0 - \frac{\omega}{2} \stackrel{{\rm o.s.}}{=} \epsilon_{k-}.  \label{os}
\end{eqnarray} 
Therefore, the contribution of a Feynman diagram, which has p-h cuts, to $\Pi_A$ is obtained by replacing the energy 
dependent vertices and T-matrices by their on-shell values. The rest is contained in $\Pi_B$. (Later we will specify 
what this means in terms of time-ordered Bethe-Goldstone diagrams, and make contact to the examples shown in Fig.4.)

A similar form of $A$ can be derived also in finite systems. In this case one uses the ``$\lambda$-representation''
\cite{MIGB} instead of the momentum representation, which involves a set of single particle 
wave functions $\Phi_{\lambda}(E)$
chosen so as to diagonalize the single particle Green function for fixed frequency $E$: $G_{\lambda' \lambda}(E)=
\delta_{\lambda' \lambda} G_{\lambda}(E)$. The form of $A$ then becomes
\begin{eqnarray}
A_{\lambda' \lambda}(k_0, \omega) = - 2\pi i \delta\left(k_0 - \frac{\epsilon_{\lambda'} + \epsilon_{\lambda}}{2}\right) 
\frac{(n_{\lambda'}-n_{\lambda}) \,Z_{\lambda'} Z_{\lambda}}{\omega - (\epsilon_{\lambda'} - \epsilon_{\lambda}) + i \gamma}
\, {\hat P}(\omega)\,, \label{al}
\end{eqnarray}
where ${\hat P}(\omega)$ implies to evaluate the function $F$, which appears in the same loop as $A$, 
at $\omega=\epsilon_{\lambda'} - \epsilon_{\lambda}$. 

Finally in this section, let us use the form (\ref{a}) to write down the expression for
$\Pi_A$ and the equation for $\Gamma$ in momentum space. For this purpose, we introduce the quasiparticle current
$j^{(\omega)}$ and the quasiparticle interaction $t^{(\omega)}$ by
\begin{eqnarray}
j^{(\omega)}(k_+,k_-) &\equiv& \sqrt{Z_{k+} Z_{k-}} \,\, \Gamma^{(\omega)}(\epsilon_{k+}\,{\bold k}_+ \, ; \,
\epsilon_{k-}\,{\bold k}_-) \label{curr} \\
t^{(\omega)}(k_+,k_-; \ell_+,\ell_-) &\equiv& \sqrt{Z_{k+} Z_{k-} Z_{\ell+} Z_{\ell-}} \,\,
T^{(\omega)}(\epsilon_{k+} {\bold k}_+,\epsilon_{k-} {\bold k}_-;
\epsilon_{\ell+} {\bold \ell}_+,\epsilon_{\ell-} {\bold \ell}_-)\,. \nonumber \\
\label{tm}
\end{eqnarray}  
Defining also the total current $j(k_+,k_-)$ in terms of the total vertex $\Gamma$ in a way similar to
(\ref{curr}), the expression for $\Pi_A$ can be written as  
\begin{eqnarray}
\Pi_A (\omega,{\bold q}) = V\,\int \frac{{\rm d}^3 k}{(2\pi)^3} {\rm tr} \left(
j^{(\omega)}(k_-,k_+) \,\frac{n_{k+}-n_{k-}}{\omega - (\epsilon_{k+} - \epsilon_{k-}) + i \gamma}
\, j(k_+,k_-) \right) \nonumber \\
\label{Piaf}
\end{eqnarray}
where $V$ is the volume of the system, and the equation for the current takes the form
\begin{eqnarray}
\lefteqn{j(k_+,k_-) = j^{(\omega)}(k_+,k_-)}  \nonumber \\
& & - \int \frac{{\rm d}^3 \ell}{(2\pi)^3} {\rm tr}\,\left[
t^{(\omega)}(k_+,k_-; \ell_+,\ell_-) \, \frac{n_{\ell +}-n_{\ell -}}{\omega - (\epsilon_{\ell +} - 
\epsilon_{\ell -}) + i \gamma} \, j(\ell_+,\ell_-)\right]\,. \nonumber \\
\label{curr1}
\end{eqnarray}
Relations analogous to (\ref{Piaf}) and (\ref{curr1}) can also be written down for finite systems
in the $\lambda$-representation.

\section{The orbital g-factor}
\setcounter{equation}{0}
In this section we will extend the previously obtained relations \cite{MIGB,BEN} for the 
orbital g-factor of a quasiparticle on the Fermi surface to the case of isospin asymmetric nuclear matter
($N \neq Z$). Although our notation will refer to the nonrelativistic case, the result is valid
also in relativistic theories.

The magnetic moment associated with the quasiparticle current ${\bold j}^{(\omega)}$ is obtained from
${\bold m}=- \frac{i}{2}\left[{\bold \nabla}_q \times \left({\bold j}^{(\omega)} 
\delta^{(3)}({\bold p}+{\bold q}-{\bold p}')\right)\right]_{q=0}$. 
In nuclear matter, the only way to get a term
proportional to the orbital angular momentum is to let the derivative act on the momentum conserving
delta function \cite{BEN}. Since the matrix element of the orbital angular momentum operator between plane wave states
is $i{\bold p}\times {\bold \nabla_p} \delta^{(3)}({\bold p}-{\bold p}')$, we obtain for the angular momentum
g-factor of a proton or neutron at the Fermi surface
\begin{eqnarray}
g_{\ell}(\alpha) = \frac{M}{e_p \cdot p_F(\alpha)} j^{(\omega)}_{\alpha} \,\,\,\,\,\,\,\,\,\,(\alpha=p,n)\,, \label{gldef}
\end{eqnarray}
where $e_p>0$ is the proton electric charge, $p_F(p)$ and $p_F(n)$ denote the Fermi momenta of protons and neutrons, 
and the current $j^{(\omega)}_{\alpha}$ refers to the $q\rightarrow 0$ limit of the
quasiparticle current (\ref{curr}) at the Fermi surface\footnote{Here and in the following sections, currents without explicit
momentum variables refer to the limit of zero momentum transfer. 
Also, in order to avoid unnecessary
indices, we will frequently use $V\equiv V^3$ and $T \equiv T^{33}$ for any vector ${\bold V}$ or
tensor $T^{ij}$ which depend on the momentum of the quasiparticle, which is put along the 3-axis.}.
The calculation of the angular momentum g-factor in nuclear
matter therefore reduces to the calculation of the quasiparticle current at the Fermi surface at zero momentum 
transfer. 

Invariance with respect to local gauge transformations of the charged particle fields leads to the Ward-Takahashi 
identity for the
electromagnetic vertex, which in turn gives expressions for the
currents $j_{\alpha}$ in the limit $\omega\rightarrow 0$ followed by $|{\bold q}|\rightarrow 0$, which is
called the ``static limit'', following ref.\cite{MIGB}. (The use of the Ward-Takahashi identity in many-body systems has
been explained in detail in refs. \cite{BEN,ASBH} for relativistic theories, but for convenience we summarize some 
relations in Appendix A.) The results for the static limit of the proton and neutron currents are
\begin{eqnarray}
j_p^{({\rm st})} = e_p\,v_F(p) \,, \,\,\,\,\,\,\,\,\,\,j_n^{({\rm st})} = 0 .  \label{static}
\end{eqnarray}     
We denote by $v_F(\alpha)$ the Fermi velocities of protons ($\alpha=p$) and neutrons ($\alpha=n$). 

Since we need the quasiparticle currents $j_{\alpha}^{(\omega)}$ instead of the static currents
$j_{\alpha}^{({\rm st})}$, we have to use eq.(\ref{curr1}) in the limit\footnote{Since the quasiparticle 
current $j^{(\omega)}$ by definition has no p-h cuts, the limits
$\omega\rightarrow 0$ followed by $|{\bold q}|\rightarrow 0$ commute for $j^{(\omega)}$. On the other
hand, the total current including p-h cuts involves the quantity $A$ of eq.(\ref{a}), which vanishes if
$|{\bold q}| \rightarrow 0$ first, but assumes a nonzero value according to eq. (\ref{fermis}) if 
$\omega \rightarrow 0$ first.}$\omega\rightarrow 0$ followed by $|{\bold q}|\rightarrow 0$.
In this limit, we can use
\begin{eqnarray}
\frac{n_{\ell +}-n_{\ell -}}{\omega - (\epsilon_{\ell +} - \epsilon_{\ell -}) + i \gamma}
\Rightarrow \frac{1}{v_F} \delta(|{\bold \ell}|-p_F)  \label{fermis} 
\end{eqnarray}
for either protons or neutrons.
Then both vectors ${\bold k}$ and ${\bold \ell}$ in eq.(\ref{curr1}) are on the Fermi surface, and the integral 
reduces to an integral over the angle between
the directions ${\hat {\bold k}}$ and ${\hat {\bold \ell}}$ (Landau angle). In this kinematics, the interaction $t^{(\omega)}$ 
depends only 
on the Landau angle and becomes the ``Landau-Migdal interaction''. It is convenient to introduce the dimensionless
interaction 
\begin{eqnarray}
{\cal F}({\hat {\bold k}}\cdot{\hat {\bold \ell}}) \equiv \frac{2 p_F^2}{\pi^2}\, \frac{1}{v_F}\, 
t^{(\omega)}({\hat {\bold k}}\cdot{\hat {\bold \ell}})\, ,  \label{deff}
\end{eqnarray}
where by definition we identify $p_F$ and $v_F$ with the Fermi momentum and the Fermi
velocity in symmetric nuclear matter with the same baryon density. It is then easy to obtain the following relations
from eq.(\ref{curr1}) and (\ref{static}) (see Appendix A):
\begin{eqnarray}
j_p^{(\omega)} &=& e_p\left[v_F(p) + \frac{v_F}{6} \left(\frac{p_F(p)}{p_F}\right)^2 F_1(pp)\right] \label{jp} \\  
j_n^{(\omega)} &=& e_p\,\frac{v_F}{6} \left(\frac{p_F(p)}{p_F}\right)^2 F_1(pn)\,, \label{jn} 
\end{eqnarray}
where $F_1(pp)$ and $F_1(pn) = F_1(np)$ are the coefficients of the $\ell=1$ Legendre polynomials 
$P_1({\hat {\bold k}}\cdot{\hat {\bold \ell}})$
in an expansion of the spin-independent part of ${\cal F}_{pp}$ or ${\cal F}_{pn}$
\footnote{The relations to the more familiar parameters $F_1$ and $F_1'$ are
$F_1(pp)=F_1+F_1'$ and $F_1(pn)=F_1-F_1'$.}. Lorentz invariance, or Galilei invariance in the nonrelativistic
case, can then be used (see Appendix A) to rewrite these expressions in the form
\begin{eqnarray}
j^{(\omega)}_{p}&=&e_p \left[\frac{p_F(p)}{\mu_p} - \frac{p_F(p) v_F}{6 p_F} F_1(pn) \frac{\mu_n}{\mu_p} 
\left(\frac{p_F(n)}{p_F}\right)^3\, \beta \right] \label{gp} \\
j^{(\omega)}_{n}&=&  e_p\,\frac{p_F(n) v_F}{6 p_F} F_1(pn)  \left(\frac{p_F(p)}{p_F}\right)^3\, \beta\,, \label{gn}
\end{eqnarray}
where $\mu_{p}$ and $\mu_n$ are the chemical potentials (Fermi energies including the rest masses)
of protons and neutrons, and
\begin{eqnarray}
\beta \equiv \frac{p_F^2}{p_F(p) p_F(n)} = \left[1-\left(\frac{N-Z}{A}\right)^2\right]^{-\frac{1}{3}}
\label{beta}
\end{eqnarray}
The orbital g-factors then can be expressed as follows
\footnote{Eqs. (\ref{gpf}), (\ref{gnf}) are exact in nuclear matter, and
include mesonic, relativistic, and configuration mixing effects. Because 
the vertex $\Gamma^{(\omega)}$ is an effective quantity in the p-h space, 
conventional RPA-type contributions, like the first order configuration mixing, 
are {\em not} included in (\ref{gpf}) and (\ref{gnf}).}  
:
\begin{eqnarray}
g_{\ell}(p)&=&\frac{M}{\mu_p} - \frac{M v_F}{3 p_F}\, F_1(pn)\, \frac{\mu_n}{\mu_p}\, \frac{N}{A}\, 
\beta \label{gpf} \\
g_{\ell}(n)&=& \frac{M v_F}{3 p_F}\, F_1(pn)\,  \frac{Z}{A}\, \beta \,. 
\label{gnf} 
\end{eqnarray}
We note that the following relation holds between the proton and neutron orbital g-factors:
\begin{eqnarray}
\frac{Z}{A} \,\frac{\mu_p}{M}\, g_{\ell}(p) + \frac{N}{A}\, \frac{\mu_n}{M}\, g_{\ell}(n) = \frac{Z}{A}\,. \label{rel}
\end{eqnarray}
In a  nonrelatistic theory we have $\mu_p=\mu_n=M$, and eq. (\ref{rel}) 
is the extension to $N\neq Z$ of the well known fact \cite{BEN,ASBH} that the isoscalar orbital g-factor is renormalized 
exclusively by relativistic effects. 

Besides the dependence on the neutron excess shown explicitly in (\ref{gpf}) and (\ref{gnf}), the Landau-Migdal
parameter $F_1(pn)$ might also depend on $(N-Z)/A$ for fixed $A$. 
If the range of the interaction is small compared to $1/p_F$, this dependence can be expected to be weak. For the case
of the long range 1-pion exchange interaction, we note that in $F_1(pn)$
the average over the cosine of the Landau angle ($\cos \Theta_L \equiv x$) is performed with respect to the 
momentum transfer in the exchange channel,
which is ${\bold Q}^2 = \left({\bold k} - {\bold \ell}\right)^2 = p_F(p)^2+p_F(n)^2-2p_F(p)p_F(n) x
=2p_F^2(1-x) + {\cal O}\left[\left(\frac{N-Z}{3A}\right)^2\right]$. Therefore, up to
${\cal O}\left(\frac{N-Z}{A}\right)$, the dependence on the neutron excess is as shown 
by the explicit factors $N/A$ and $Z/A$ in the relations (\ref{gpf}) and (\ref{gnf}).

\section{The E1 sum rule and the $\kappa-g_{\ell}$ relation}
\setcounter{equation}{0}

While the treatment of the orbital g-factor in the previous section was completely general,
the discussion of the E1 sum rule is plagued by problems related to the spurious
center of mass motion \cite{EG}. Therefore, from now we have to restrict ourselves to a
nonrelativistic framework, and consider the case of a Hamiltonian 
\begin{eqnarray}
{\hat H} &=& \frac{-1}{2M} \int {\rm d}^3 r\,\, \psi^{\dagger}({\bold r})\, \Delta \, \psi({\bold r})  \label{h0} \\
&+& \frac{1}{2} \int {\rm d}^3 r_1 \int {\rm d}^3 r_2 \,\, \psi^{\dagger}({\bold r}_1)\psi^{\dagger}({\bold r}_2)
\,V({\bold r}_1,{\bold r}_2)\, \psi({\bold r}_2)\psi({\bold r}_1) \label{h1} \\
&\equiv& {\hat H}_0 + {\hat V} \label{defh}
\end{eqnarray}
with the two-body potential
\begin{eqnarray}
V({\bold r}_1,{\bold r}_2)=V_0({\bold r}_1,{\bold r}_2) + {\bold \tau}_1 \cdot {\bold \tau}_2 \,
V_1({\bold r}_1,{\bold r}_2). \label{pot}
\end{eqnarray}

\subsection{Effective charges and gauge invariance}
In this subsection, we digress to a discussion of local gauge invariance of a Hamiltonian containing two-body potentials.
Our aims are, first, to investigate whether the introduction of the familiar E1 effective charges, which
eliminate the spurious center of mass motion, is consistent with gauge invariance, and, second, derive some
relations for the 2-body part of the electromagnetic interaction which will be used later.

Let us consider local gauge transformations of the form
\begin{eqnarray}
\psi({\bold r}) \rightarrow {\rm exp}\left[i  {\hat Q} \chi({\bold r})\right]\, \psi({\bold r})
\label{gauge}
\end{eqnarray}
where $\chi({\bold r})$ is the gauge function, and
\begin{eqnarray}
{\hat Q} \equiv \frac{1+\tau_z}{2}\, Q_p + \frac{1-\tau_z}{2}\, Q_n\, .  
\label{iso}
\end{eqnarray}
In the first place, of course, we take the physical electric charges 
$e_p=-e >0$ and $e_n=0$ for $Q_p$ and $Q_n$,
but we wish to keep our expressions more general in order to incorporate the effective charges in the later discussions.
This gauge transformation effectively changes the first quantized operators as follows:
\begin{eqnarray}
H_0(i) &\rightarrow& {\rm exp}\left[-i {\hat Q}^{(i)} \chi_i \right] H_0(i) \left[i {\hat Q}^{(i)} \chi_i \right] 
\label{th0} \\
V(i,j) &\rightarrow& {\rm exp}\left[-i\left({\hat Q}^{(i)} \chi_i + {\hat Q}^{(j)} \chi_j \right)\right] V(i,j)
\left[i\left({\hat Q}^{(i)} \chi_i + {\hat Q}^{(j)} \chi_j \right)\right],  \nonumber \\
\label{th1}
\end{eqnarray}
where $\chi_i\equiv \chi({\bold r}_i)$, $\chi_j\equiv \chi({\bold r}_j)$, and the other notations are
self evident. Although it is not difficult to work out these expressions for a general gauge function,
we consider here the long wave length limit (LWL) limit in order to derive the necessary relations most
directly. In this case it is sufficient\footnote{This is easily seen \cite{FUI} by noting that the vector potential
in the LWL is a constant, and a gauge transformation ${\bold A}\rightarrow {\bold A}
+{\bold \nabla}\chi$ should again give a field of the same kind.} to consider $\chi({\bold r}) = c_0 + {\bold r}\cdot {\bold c}$, 
where the $c_{\alpha}$ are constants. Using the fact that the 2-body potential commutes with the total
2-body isospin operator, the quantity $\left({\hat Q}^{(i)} \chi_i + {\hat Q}^{(j)} \chi_j \right)$ in the gauge transformation
(\ref{th1}) can be replaced by ${\bold c}\cdot \left({\bold D}_Q^{(i)} + {\bold D}_Q^{(j)}\right)$, 
where ${\bold D}_Q^{(i)}={\hat Q}^{(i)}{\bold r}_i$
is the dipole operator for particle $i$, referring to arbitrary charges $Q_p$ and $Q_n$. 
Since also (\ref{th0}) can be expressed
by the dipole operator for particle $i$, the gauge transformation of the total Hamiltonian in first quantization
\begin{eqnarray}
H = H_0 + V \equiv \sum_{i} H_0(i) + \sum_{i<j} V(i,j)  \label{ham1}
\end{eqnarray}
can be written in terms of the dipole operator ${\displaystyle {\bold D}_Q=\sum_i {\bold D}_Q^{(i)}}$ as follows:
\begin{eqnarray}
H \rightarrow {\rm exp}\left[-i {\bold c}\cdot {\bold D}_Q\right]\, H \, {\rm exp}\left[i {\bold c}\cdot 
{\bold D}_Q\right]\,. \label{dip}
\end{eqnarray}
This expression can be expanded in a series of multi-commutators:
\begin{eqnarray}
H \rightarrow H -i {\bold c}\cdot \left[{\bold D}_Q,H\right] - \frac{1}{2} c^i c^j \left[D^i_Q,\left[D^j_Q,H
\right]\right] + \dots \label{haus}
\end{eqnarray}
where we have shown the terms up to ${\cal O}(e^2)$ explicitly. The Hamiltonian can then be made locally
gauge invariant up to some power of $e$ by introducing the gauge (photon) field
${\bold A}$ which transforms under the gauge transformation as ${\bold A}\rightarrow {\bold A}+{\bold \nabla}\chi$.
It is then straight forward (see Appendix B) to show that the following Hamiltonian is locally gauge invariant 
up to ${\cal O}(e^2)$:
\begin{eqnarray}
{\cal H}(Q_p,Q_n) &=& H + i {\bold A}\cdot \left[{\bold D}_Q,H\right] - \frac{1}{2} A^i A^j \left[D^i_Q,
\left[D^j_Q,H\right]\right] \label{comm} \\
&\equiv& H + \delta H_0(Q_p,Q_n) + \delta V(Q_p,Q_n),
\end{eqnarray}
where $\delta H_0$ ($\delta V$) is the contribution from $H_0$ ($V$) to the commutator terms. 
Evaluating the part $\delta H_0$ explicitly, we arrive at the familiar form 
\begin{eqnarray}
{\cal H}(Q_p,Q_n) = \frac{1}{2M} \sum_{i=1}^Z \left({\bold p}_i - Q_p {\bold A} \right)^2
+ \frac{1}{2M} \sum_{i=1}^N \left({\bold p}_i - Q_n {\bold A} \right)^2 + \delta V(Q_p,Q_n) \,, \nonumber \\
\label{hel}
\end{eqnarray}
where 
\begin{eqnarray}
\delta V(Q_p,Q_n) = i {\bold A}\cdot \left[{\bold D}_Q,V\right] - \frac{1}{2} A^i A^j \left[D_Q^i,
\left[D_Q^j,V\right]\right]\,. \label{commv}
\end{eqnarray}
The second term on the r.h.s. of (\ref{comm}) involves the familiar form of the current operator in the LWL: 
${\bold j}_Q({\bold q}=0)=i[H,{\bold D}_Q]$. The third term on the r.h.s. involves the 2-body part of the Compton scattering
amplitude in the LWL, as will be discussed in more detail in sect. 6.

Here we note the following point concerning the dependence of the interaction term $\delta V$ on the
charges $Q_p$ and $Q_n$: If we consider a pair of nucleons denoted by $(1,2)$, the relevant term in the
dipole operator is
\begin{eqnarray}
{\bold D}_{Q,12} = {\hat Q}^{(1)} {\bold r}_1 + {\hat Q}^{(2)} {\bold r}_2  = {\bold R} \left(Q + T_z\,\Delta Q\right)
+ \frac{\bold r}{4}\,\Delta Q \, \left(\tau_z^{(1)} - \tau_z^{(1)}\right), \label{dipole}
\end{eqnarray}
where ${\bold R}$ and ${\bold r}$ are the c.m. and relative coordinates of the pair $(1,2)$, 
$T_z = \left(\tau_z^{(1)}+ \tau_z^{(1)}\right)/2$, and $Q \equiv Q_p+Q_n$, $\Delta Q \equiv Q_p - Q_n$. 
Using this expression in (\ref{commv}), we see that, if the potential commutes with the 
2-body c.m. coordinate ($\left[V,{\bold R} \right] = 0$),
the 2-body current (first term on the r.h.s. of eq.(\ref{commv})) depends only on the difference
of the charges $\Delta Q$ and not on the sum $Q$. From the Jacobi identity for double commutators it is seen
that the same holds also for the two-body part of the photon scattering amplitude (second term on the r.h.s.
of eq.(\ref{commv}))
\footnote{This is seen as follows: First, if we apply the Jacobi identity to the double commutator $[T_z,[D_Q^i,V]]$, 
we obtain $[T_z,[D_Q^i,V]]=0$. Second, if we apply it to $[R^i,[V,D_Q^j]]$, we obtain $[R^i,[V,D_Q^j]]=0$ if $V$
commutes with $R^i$. From these two observations, it follows that in the double commutator
$[D_Q^i,[D_Q^j,V]]$ the term proportional to the c.m. coordinate in the dipole operator (\ref{dipole}) 
does not contribute, and therefore this double commutator depends only of the difference $\Delta Q$ of the charges.}. 
This is important for our later discussions, since the effective charges for E1 
transitions, $q_p=e_p\,N/A$, $q_n=-e_p\,Z/A$, satisfy $\Delta q= e_p$ as do the physical electric charges. 
Therefore, if the two-body potential commutes with the two-body c.m. coordinate, the interaction terms in the 
Hamiltonian (\ref{hel}) are invariant under the replacement of the physical charges are by the effective ones:
$\delta V(q_p,q_n) = \delta V(e_p,e_n)$. 

The introduction of the effective charges is then done as usual: First, for the physical charges we have
\begin{eqnarray}
{\cal H}(e_p,e_n) = H_0 - \frac{e_p}{M} \sum_{i=1}^Z {\bold p}_i \cdot {\bold A} + \frac{e_p^2}{2M} 
\sum_{i=1}^Z {\bold A}^2 + \delta V(e_p,e_n) \,.  \label{hem0}
\end{eqnarray}
Second, in order to remove the
effect of the center of mass motion on the electromagnetic interaction, we replace the 
momenta ${\bold p}_i$ in the second term of (\ref{hem0}) by the
Jacobi momenta ${\bold p}_i - {\bold P}/A$, where 
${\bold P}=\sum_i {\bold p}_i$ is the total momentum \cite{COM1}. This replacement naturally leads to the effective charges
\begin{eqnarray}
q_p=\frac{N}{A}\,e_p\,, \,\,\,\,\,\,\,\,\,\,\,\,q_n=-\frac{Z}{A}\,e_p\,.  \label{effc}
\end{eqnarray}
Third, since ${\bold A}$ is a constant in the LWL considered here, we can use the following identity for the third term on the 
r.h.s. of (\ref{hem0}):
\begin{eqnarray}
Z e_p^2 = \frac{Z^2 e_p^2}{A} + Z q_p^2 + N q_n^2  \label{iden}
\end{eqnarray}
to obtain the final form of the Hamiltonian as
\begin{eqnarray}
{\hat {\cal H}}(q_p,q_n) = \frac{Z^2 e_p^2}{2 M A}\, {\bold A}^2 + {\cal H}(q_p,q_n).  \label{new0}
\end{eqnarray}
Here the first term describes the Thomson scattering of the nucleus as a whole \cite{EG}, and ${\cal H}(q_p,q_n)$ is 
given by (\ref{hel}) with the effective charges (\ref{effc}).

The procedure to remove the c.m. motion therefore leads to the Hamiltonian ${\cal H}(q_p,q_n)$, which is 
locally gauge invariant under the gauge transformation involving
the effective charges. Therefore one can continue to use all consequences of local gauge
invariance, like Ward identities etc., for the theory described by the Hamiltonian ${\cal H}(q_p,q_n)$.

\subsection{The E1 sum rule}
The strength
function (cross section) for the absorption of unpolarized photons by a nucleus in its ground state
$|0\rangle$ is given by \cite{EG}
\begin{eqnarray}
S(\omega)&=&4\pi^2 \sum_n |\langle n| {\hat j}({\bold q}) \frac{1}{\sqrt{\omega}}|0\rangle|^2
\,\delta(\omega-\omega_{n0}) \label{str1}   \\
&\stackrel{\rm LWL}{\longrightarrow}&
4\pi^2 \sum_n |\langle n| {\hat j}({\bold q}=0) \frac{1}{\sqrt{\omega}}|0\rangle|^2
\,\delta(\omega-\omega_{n0}), \label{lwl} 
\end{eqnarray}  
where ${\hat j}({\bold q})$ denotes the Fourier transform of the 3-component (or any other space component) 
of the electromagnetic current operator 
\footnote{We mark the current operators and quasiparticle currents for the E1 effective
charges by hats in order to distinguish them from the quantities related to the physical charges.} 
for the effective charges (\ref{effc}), and $\omega_{n0}=E_n-E_0$ is the
excitation energy of the state $|n\rangle$. The factor $\frac{1}{\sqrt{\omega}}$ in (\ref{str1}) and (\ref{lwl}) 
originates from the normalization of the photon field. The LWL indicated in (\ref{lwl})
holds if $|{\bold q}|R<<1$, where $R$ is the nuclear radius\footnote{The multipole expansion of (\ref{str1}) can be performed
as usual by the expansion of the factor ${\rm exp}(-i {\bold q}\cdot {\bold r})$ which defines the Fourier transform
of the current operator. This leads to the J-multipole operators of the
form $\left[j^{(1)}\times Y^{(l)}\right]^{(J)} j_l(qr)$. The E1 multipole ($J=1$) therefore contains, besides the
LWL, also ``retardation terms'' $\propto (qR)^2$, which come from the expansion of the Bessel functions for $l=0$ and 
$l=2$. Therefore, the E1 multipole contains also terms which go beyond the LWL.}.

In the LWL one can use current conservation to express the current operator as
${\hat j}({\bold q}=0)=i[H,D_q]$, where $D_q$ is the 3-component of the 
dipole operator involving the effective charges (\ref{effc}). Then one arrives at the ``nonretarded'' E1 sum rule 
\cite{EG} 
\begin{eqnarray}
S\equiv\int_0^{\infty} {\rm d}\omega\, S(\omega) &=& 4\pi^2 \sum_n \frac{1}{\omega_{n0}} 
\,|\langle n| {\hat j}({\bold q}=0) |0\rangle|^2 \label{sum1} \\
&=& -2 \pi^2 \langle 0|\left[\left[H,D_q \right],D_q \right]|0\rangle.
\label{sum2}
\end{eqnarray}
Of course, this sum rule is not directly observable, since it includes only the unretarded
E1 multipole, and for high $\omega$ ($=|{\bold q}|$ for a physical photon) the LWL is not valid. It is therefore important 
to identify a part of the sum rule which holds in the low energy region where the LWL is justified. 

We now relate the strength function and sum rule to the Fourier transform of the correlation function (see eq. (\ref{pol}))
\begin{equation}
\Pi^{\mu \nu}({\bold q}',{\bold q};\omega)=i \int_{-\infty}^{\infty} {\rm d}\tau\,
e^{i\omega \tau} 
\langle 0|T\left({\hat j}^{\mu}({\bold q}',t') {\hat j}^{\nu}(-{\bold q},t) \right)|0\rangle,
\label{corr}
\end{equation}
where $\tau=t'-t\,$, and ${\bold q}$ and ${\bold q}'$ are the incoming and outgoing momenta.
For the case of forward scattering we define
$\Pi^{\mu \nu}({\bold q},\omega)\equiv \Pi^{\mu \nu}({\bold q},{\bold q};\omega)$, and 
obtain from eq.(\ref{corr}) the spectral representation for the 33-component
\begin{equation}
\Pi({\bold q},\omega)=-\sum_n |\langle 0|{\hat j}({\bold q})|n\rangle |^2
\left(\frac{1}{\omega-\omega_{n0}+i\delta}-\frac{1}{\omega+\omega_{n0}-i\delta}\right).
\label{kl}
\end{equation}
Comparison with eqs. (\ref{str1}) and (\ref{sum1}) gives
\begin{eqnarray}
S(\omega)&=&\frac{4\pi}{\omega} \,\,{\rm Im} \,\Pi({\bold q}=0,\omega) 
\,\,\,\,\,\,\,\,\,\,
\,\,\,\,\,(\omega>0) \label{stra} \\
S &=& 2\pi^2 \,\,\Pi({\bold q}=0,\omega=0). \label{suma}
\end{eqnarray}
So far, all relations of this subsection hold for finite nuclei. The strength function in the LWL,
eq. (\ref{stra}), is actually meaningless in nuclear matter, since there is no scale which 
corresponds to the nuclear radius $R$. The sum rule (\ref{suma}), however, represents a bulk property
of the nucleus which should not depend on the details of nuclear structure, and therefore the discussion
of the sum rule in nuclear matter has been the subject of many previous investigations
\cite{FUI}-\cite{FIH}.     

In order to apply eq.(\ref{suma}) in nuclear matter and to discuss the
$\kappa$-$g_{\ell}$ relation, we have to know the order in which the limits 
${\bold q}\rightarrow 0,\,\,\omega\rightarrow 0$ should be taken. This is important,
because if ${\bold q}\rightarrow 0$ is taken first (the ``$\omega$-limit'' of sect. 3), the p-h excitations 
do not contribute to the polarization, while for $\omega\rightarrow 0$ first (the
``static limit'') 
they contribute at the Fermi surface \cite{BEN}. We therefore present an alternative derivation of 
eq.(\ref{suma}), which makes clear the order of the limits. We make use of the following identity \cite{COM2}:
\begin{equation}
\frac{\partial}{\partial x'^{\mu}}\,\frac{\partial}{\partial x^{\nu}} 
\langle 0|T\left({\hat j}^{\mu}(x')\,{\hat j}^{\nu}(x)\right)|0\rangle =
- \langle 0|\left[{\hat j}^0(x'),{\hat j}^0(x)\right]|0\rangle \left(\frac{\partial}{\partial \tau}\,\,
\delta\left({\tau}\right)\right).
\label{ide}
\end{equation}
The use of a partial integration in $\tau$ and the Heisenberg equation of motion for ${\hat j}^0$ 
then gives the following Ward-Takahashi identity for the correlation function (\ref{corr}):
\begin{equation}
q'_{\mu} q_{\nu} \Pi^{\mu \nu}({\bold q}',{\bold q};\omega)=
- \langle 0|\left[\left[H,{\hat j}^0(-{\bold q})\right],{\hat j}^0({\bold q}')\right]|0\rangle
+\omega \,\,\langle 0|\left[{\hat j}^0(-{\bold q}),{\hat j}^0({\bold q}')\right] |0\rangle
\label{wt}
\end{equation}
where $q'^{\mu}=(\omega,{\bold q}')$ and $q^{\mu}=(\omega,{\bold q})$. We set $\omega=0$
in this identity, then apply ${\displaystyle \frac{\partial}{\partial q'_i}\,\frac{\partial}{\partial q_j}}$
to both sides, and finally let ${\bold q'}$ and ${\bold q}$ go to zero to obtain the 
low-energy theorem 
\begin{equation}
\Pi({\bold q}=0, \omega=0)=- \langle 0|\left[\left[H,D_q\right],D_q \right]|0\rangle,
\label{let}
\end{equation}
where now the limit is defined as $\omega\rightarrow 0$ first followed by ${\bold q}\rightarrow 0$
(``static limit'').
Comparison with eq. (\ref{sum2}) then gives again the relation (\ref{suma}) with this
particular ``static limit'' prescription for the low frequency\,/\,low wave length limit. 

Comparison with eq.(\ref{comm}) of the previous section shows that the double commutator in (\ref{let}), 
which determines the sum rule (\ref{sum2}), is the ``seagull part'' of the low energy photon scattering amplitude.  
This point will be further discussed in sect. 6.

\subsection{The $\kappa-g_{\ell}$ relation}
 
The polarization, which determines the sum rule by (\ref{suma}), has been split into two
parts in eq.(\ref{res}). We will now show that the part $\Pi_A$, which has been given explicitly 
in (\ref{Piaf}), is related to the orbital g-factors. The part $\Pi_B$, on the other hand, has no relation
to the g-factors.

Taking the static limit ($\omega\rightarrow 0$ first, then ${\bold q}\rightarrow 0$) of $\Pi_A$ in (\ref{Piaf}),
we obtain
\begin{eqnarray}
\Pi_A(\omega=0, {\bold q}=0) = {\hat j}^{(\omega)}_p \frac{Z}{v_F(p) p_F(p)} {\hat j}_p^{({\rm st})}
+  {\hat j}^{(\omega)}_n \frac{N}{v_F(n) p_F(n)} {\hat j}_n^{({\rm st})}\,,  \label{pia0}
\end{eqnarray}
where the currents ${\hat j}^{(\omega)}$ and ${\hat j}^{({\rm st})}$ at the Fermi surface refer to the 
{\em effective} charges, in contrast to the currents for the physical charges used in sect.3.

We now follow the same steps as in sect. 3: The Ward identities determine the
currents in the static limit as 
\begin{eqnarray}
{\hat j}_p^{({\rm st})} = q_p v_F(p) \,, \,\,\,\,\,\,\,\,\,\,{\hat j}_n^{({\rm st})} = q_n v_F(n) .  \label{static1}
\end{eqnarray}     
Relation (\ref{curr1}) then gives the currents in the $\omega$-limit:
\begin{eqnarray}
{\hat j}_p^{(\omega)} &=& q_p \left[v_F(p) + \frac{v_F}{6} \left(\frac{p_F(p)}{p_F}\right)^2 F_1(pp)\right] 
+ q_n \frac{v_F}{6} \left(\frac{p_F(n)}{p_F}\right)^2 F_1(pn) \nonumber \\
\label{jp1} \\  
{\hat j}_n^{(\omega)} &=& q_n \left[v_F(n) + \frac{v_F}{6} \left(\frac{p_F(p)}{p_F}\right)^2 F_1(nn)\right]
+ q_p \frac{v_F}{6} \left(\frac{p_F(p)}{p_F}\right)^2 F_1(pn)\,. \nonumber \\
\label{jn1}   
\end{eqnarray}
If we use the Landau effective mass relations of Appendix A for the nonrelativistic case, 
we obtain the very simple results
\begin{eqnarray}
{\hat j}_p^{(\omega)} = j_p^{(\omega)} - Z e_p\,\frac{p_F(p)}{A M} \,, 
\,\,\,\,\,\,\,\,\,\,\,\,\,\,\,\,\,\,\,\,
{\hat j}_n^{(\omega)} = j_n^{(\omega)} - Z e_p\,\frac{p_F(n)}{A M}\,, \label{hat} 
\end{eqnarray}
where the currents $j_p^{(\omega)}$ and $j_n^{(\omega)}$ for the physical charges are given
by eqs. (\ref{gp}) and (\ref{gn}) in the nonrelativistic limit $\mu_p=\mu_n=M$.
The relations (\ref{hat}) clearly show the role of the effective
charges to remove the c.m. motion: For the case $N=Z$ the subtractions in 
(\ref{hat}) make sure that there are no contributions from isoscalar currents to the 
collective motion of the system. 

Inserting the static currents (\ref{static1}) and the $\omega$-currents (\ref{hat}) into (\ref{pia0}), and 
using the definition of the orbital g-factors (\ref{gldef}), we obtain
\begin{eqnarray}
\Pi_A(\omega=0, {\bold q}=0) = e_p^2 \frac{NZ}{AM} \left(g_{\ell}(p) - g_{\ell}(n)\right). \label{erg}
\end{eqnarray}
The sum rule (\ref{suma}) then can be expressed as
\begin{equation}
S=({\rm TRK})\left(1+\kappa_A+\kappa_B\right) \equiv S_A + S_B\,,   \label{trk}
\end{equation}
where ${\displaystyle ({\rm TRK})=2 \pi^2 e_p^2  \frac{NZ}{AM}}$ is the Thomas-Reiche-Kuhn sum rule value, 
and the enhancement factor $1+\kappa$ has been split into the two parts
$1+\kappa_A$ and $\kappa_B$ originating from $\Pi_A$ and $\Pi_B$, respectively:
\begin{eqnarray}
1+\kappa_A &=& \frac{1}{e_p^2} \frac{AM}{NZ}\, \Pi_A({\bold q}=0,\omega=0) =  
g_{\ell}(p) - g_{\ell}(n) \label{fh} \\
\kappa_B &=& \frac{1}{e_p^2} \frac{AM}{NZ}\, \Pi_B({\bold q}=0,\omega=0).      \label{kb}
\end{eqnarray}
Using (\ref{rel}), the relation (\ref{fh}) between $\kappa_A$ and the orbital 
g-factors can also be expressed as follows:
\begin{eqnarray}
g_{\ell}(p) &=& 1 + \frac{N}{A} \kappa_A \label{prot} \\
g_{\ell}(n) &=& -  \frac{Z}{A} \kappa_A\,, \label{neut} 
\end{eqnarray}
where $\kappa_A$ is given by
\begin{eqnarray}
\kappa_A = - \frac{M v_F}{3\,p_F} \, F_1(pn)\,\left[1-\left(\frac{N-Z}{A}\right)^2\right]^{-\frac{1}{3}}.
\label{kap}
\end{eqnarray}
From our discussions in sect. 3 concerning the dependence of $F_1(pn)$ on the neutron excess, we see that
the deviation of $\kappa_A$ from its value for symmetric nuclear matter starts with 
$\left((N-Z)/A\right)^2$.

\section{Discussion}
\setcounter{equation}{0}

As we pointed out in sect.3, the expressions (\ref{gpf}) and (\ref{gnf}) are exact in nuclear
matter and therefore take into account all possible contributions from meson exchange currents and
configuration mixings, except for the conventional RPA-type contributions (e.g., the first order configuration
mixing). The E1 enhancement factor $\kappa$, on the other hand, has been split
into the pieces $\kappa_A$ and $\kappa_B$ in eq.(\ref{trk}), and only $\kappa_A$ is related to the orbital
g-factors. In this section we wish to discuss which processes are taken into account by $\kappa_A$,
and whether $\kappa_A$ can be related to the experimentally measured sum rule.

For this purpose, we return to eq.(\ref{res}), where the polarization has been split into the two
pieces $\Pi_A$ and $\Pi_B$. According to the optical theorem (\ref{stra}), the strength function
for finite nuclei in the LWL then splits into two pieces as well: 
$S(\omega)=S_A(\omega)+S_B(\omega)$. We first wish to discuss the
piece $\Pi_A$ in terms of Bethe-Goldstone diagrams. From the structure of eqs. (\ref{gfu}), (\ref{res}), 
and the energy denominator appearing in the quantity $A$ of (\ref{al}), it is clear that the piece 
$\Pi_A$ can be related only to those
time-ordered diagrams which have a p-h cut in each internal loop integral appearing in the RPA series,
i.e., which can be made disconnected by cutting a p-h pair in any internal loop of the RPA series. 
Diagrams without p-h cuts contribute
exclusively to $\Pi_B$, see Fig. 5 for examples. 

Let us denote the contribution of a particular time-ordered diagram 
$i$, which involves
p-h cuts, to the total polarization by $\Pi^{(i)}$, see Fig. 4 for examples.   
Besides the p-h cuts, which correspond to ``low energy denominators'' of the form 
$\left(\omega-E^{\rm ph}\pm i \delta\right)^{-1}$, 
the diagram $i$ will in general also involve ``high energy denominators'' 
$\left(\omega-E^{\rm high} + i \delta\right)^{-1}$. Here $E^{\rm ph}$ denotes the excitation energy of an intermediate
p-h pair, and $E^{\rm high}$ is the excitation energy of a more complicated intermediate state like a 2p-2h state etc. 
These high energy denominators appear in the effective vertex $\Gamma^{(\omega)}$ (second row in Fig. 4)
and in the 
effective interaction $T^{(\omega)}$ (third row of Fig. 4). The diagram $i$ will therefore in general give 
contributions to both $\Pi_A$ and $\Pi_B$, that is, $\Pi^{(i)} = \Pi_A^{(i)}+\Pi_B^{(i)}$. 
In Appendix C we show that the piece 
$\Pi_A^{(i)}$ can be obtained from the full $\Pi^{(i)}$ by approximating the high energy denominators 
$\left(\omega-E^{\rm high} + i \delta\right)^{-1}$
as follows:
\begin{enumerate}
\item In the energy denominators which come from the effective vertex $\Gamma^{(\omega)}$, the replacement
$\omega \rightarrow E^{\rm ph}$ is introduced. That is,
$\omega$ is replaced by the excitation energy of the p-h pair which enters (or leaves) the vertex
$\Gamma^{(\omega)}$ in the loop under consideration.
\item In the energy denominators which come from the effective interaction $T^{(\omega)}$, the replacement
$\omega \rightarrow \frac{1}{2} (E_i^{\rm ph} + E_f^{\rm ph})$
is introduced. That is, $\omega$ is replaced by the average of the
excitation energies of the p-h pairs which enter ($E_i^{\rm ph}$) and leave ($E_i^{\rm ph}$) the block 
$T^{(\omega)}$.
\item If the particle or the hole in the intermediate state under consideration (excitation energy $E^{\rm ph}$) has 
an energy dependent (dispersive) 
self energy, the associated high energy denominators are replaced by the first two terms of their 
expansion around $\omega=E^{\rm ph}$. That is, the effects of dispersive self energies are included only in 
the quasiparticle energies and residues.
\end{enumerate}
To summarize these prescriptions: In order to obtain the contribution of a time-ordered graph, 
which has a p-h cut, to $\Pi_A$, one has to replace $\omega$ in the high energy denominators
by a value which is related to the excitation energy of the intermediate p-h pair. 

Let us then consider the contribution of a particular time-ordered graph, $\Pi^{(i)}$, which has both 
p-h and higher energy cuts, to the LWL sum rule. In Appendix C we show from analyticity that the following
relation\footnote{As long as we consider a particular diagram $i$, the imaginary part ${\rm Im}\, \Pi^{(i)}$ is not
necessarily positive.} holds for any diagram $i$:
\begin{eqnarray}
S^{(i)} \equiv 4 \pi \int_0^{\infty} \frac{{\rm d} \omega}{\omega}\, {\rm Im}\, \Pi^{(i)}({\bold q}=0, \omega) 
= 2 \pi^2 \, \Pi ^{(i)}({\bold q}=0, \omega=0). \label{dis}
\end{eqnarray}
Since the analyticity arguments leading to (\ref{dis}) are not invalidated by the replacement
of the high energy denominators by $\omega$-independent quantities following the lines discussed above, 
a relation similar to (\ref{dis}) holds
for $\Pi_A^{(i)}$ (and therefore also for $\Pi_B^{(i)}$) separately:
\begin{eqnarray}
S_A^{(i)} \equiv 4 \pi \int_0^{\infty} \frac{{\rm d} \omega}{\omega}\, {\rm Im}\, \Pi_A^{(i)}({\bold q}=0, \omega) 
= 2 \pi^2 \, \Pi_A ^{(i)}({\bold q}=0, \omega=0). \label{disa}
\end{eqnarray}
On the other hand, by using the above prescriptions to obtain $\Pi^{(i)}_A$ from the full $\Pi^{(i)}$, we
see that, as long as the energies $E^{\rm high}$ are on the average large
compared to the typical p-h excitation energies $E^{\rm ph}$, the following relation holds: 
\begin{eqnarray}
\Pi^{(i)}_A({\bold q}=0, \omega=0) \simeq \Pi^{(i)}({\bold q}=0, \omega=0).  \label{appr}
\end{eqnarray}
Here the symbol $\simeq$ in (\ref{appr}) indicates that this relation is approximately valid in finite
nuclei as long as in the average $E^{\rm high}>>E^{\rm ph}$, while in nuclear matter it is exact because
for ${\bold q}=0$ we have $E^{\rm ph}=0$ from momentum conservation.
Combining with eq.(\ref{dis}) we obtain the following relation for the contribution of the diagram $i$
to the sum rule: 
\begin{eqnarray}
S^{(i)} \simeq S_A^{(i)}. \label{di} 
\end{eqnarray}
This relation indicates that the higher excited states (2p-2h etc.), which are mixed into the low energy
p-h states in the diagram $i$ under consideration, contribute to the sum rule mainly via their real
parts. Their imaginary parts lead to deviations of the sum rule $S^{(i)}$ from the value $S_A^{(i)}$, but these
deviations are small, although the strength function ($S^{(i)}(\omega)$) itself is, off course, influenced
by these imaginary parts, which give rise to the well known ``spreading widths''.  

Summing over all diagrams $i$ which have p-h cuts and possibly also higher energy cuts, we then have the following 
relation for their contribution to the sum rule: 
\begin{eqnarray}
\sum_i \, S^{(i)} \simeq S_A = ({\rm TRK}) \left(g_{\ell}(p)-g_{\ell}(n)\right),  \label{conc}
\end{eqnarray}
Our above discussion indicates that, although in principle the $\kappa-g_{\ell}$ relation takes into account 
the mixing of the higher excited states (2p-2h etc.) into the p-h states only via their real parts,
the contribution of the imaginary parts of the higher excited states to the sum rule is comparatively small.
Therefore we can say that, besides the contributions from the p-h cuts, the $\kappa-g_{\ell}$ relation includes also the most 
important effects of the mixing between the p-h and the higher excited states.
In the same approximation, the remaining piece $\kappa_B$
arises exclusively from those diagrams which have no p-h cuts at all, like those shown in Fig. 5.

Since the part $\kappa_A$ includes the p-h cuts and the most important part of the coupling to higher
excited states, one can expect that $\kappa_A$ will be dominated by the GDR, which is
a superposition of collective p-h pairs. We can understand this from  
eq.(\ref{res}), which shows that the part $\Pi_A$ involves the total vertex $\Gamma$, which is obtained from the effective
p-h vertex $\Gamma^{(\omega)}$ via the RPA-type equation (\ref{gfu}). The vertex $\Gamma$ is thus
generated by a collective superposition of p-h states represented by the p-h
propagator $A$, and contains the discrete poles as well as the continuum cut due to the collective
p-h pairs. The discrete poles correspond to the ``zero sound'' modes in nuclear matter, and to the
giant resonances in finite nuclei. Therefore, the piece $\kappa_A$ includes the effect of the most prominent
excitations, that is the collective p-h excitations, in the low energy region, and also the most important
part of their mixing to the higher excited states. 

On the other hand, the piece $\kappa_B$ arises from those diagrams which have no p-h
cuts, like the examples shown in Fig. 5. It is known that, because of the short range nature of the tensor
force between the nucleons, the ``effective energy denominators'' for these diagrams are 
large, typically several hundred MeV \cite{ABHI,IHB}. These diagrams will therefore contribute to the strength
function mainly in the region well beyond the GDR. In this high energy region, the
LWL is no longer valid, that is, the contribution of these diagrams to the LWL sum rule (\ref{suma})
has no connection to the measured strength function. The 
$\kappa-g_{\ell}$ relation (\ref{fh}) therefore connects the ``observable'' part of the enhancement factor in the
LWL sum rule, which is $1+\kappa_A$, to other observable quantities, namely the angular momentum g-factors. 

The important point of our above discussions was the observation that the part $\Pi_A$ includes 
the effects which come from the collective p-h excitations.  
In Appendix D we illustrate this point more explicitly for the case of nuclear matter.
There we show that the RPA-type equation for the vertex (\ref{gfu}) is equivalent to the
Landau equation \cite{MIGB,NOZ,TBA} in the vector-isovector channel, and derive the expression for 
$\Pi_A$ in terms of the solutions to the Landau equation, see eq.(\ref{re2}). 
The discussions in Appendix D clearly demonstrate that the effects of the collective 
excitations of the system are included in the part $\Pi_A$.    

Let us come back to the $\kappa-g_{\ell}$ relation (\ref{fh}), and discuss the nuclei in the $^{208}$Pb region
for the sake of illustration. The theoretical values of $g_{\ell}(p)$ and
$g_{\ell}(n)$ for nuclei in the lead region given in table 7.12 of ref. \cite{ASBH} are
$g_{\ell}(p) \simeq 1+0.13$, $g_{\ell}(n) \simeq -0.07$. These values are consistent
with (\ref{rel}), and with the empirical ones of ref. \cite{YAM}. From Eq.(\ref{fh}) one then 
obtains the estimate $\kappa_A \simeq 0.2$. In the analysis of
ref.\cite{DAL}, which uses the experimentally measured scattering cross section to extract
the total photoabsorption cross section via dispersion relations, it was concluded
that ``any reasonable prescription gives (experimental) values of $\kappa_{\rm GDR}$ 
between 0.2 and 0.3'', where $\kappa_{\rm GDR}$ was extracted from the area under a Lorentzian curve fitted to the GDR.  
This would indicate at least a qualitative consistency between theory and experiment, since our $\kappa_A$ 
can be identified with $\kappa_{\rm GDR}$ as discussed above. However, the analysis of Ref.\cite{SCHU2} 
indicates that $\kappa_{\rm GDR}$ might be enhanced by retardation corrections.
This would indicate a discrepancy between theory and experiment, which should be further investigated 
\footnote{Since in the discussion of refs.\cite{SCHU2} and \cite{DAL}   
(and also in other papers) it was assumed the quantity $g_{\ell}$, which enters in the $\kappa-g_{\ell}$ 
relation, includes only the meson exchange current corrections to the orbital g-factor and not the other
nuclear structure effects (configuration mixings etc), it was concluded that theory would 
be consistent with the larger value for $\kappa_{\rm GDR}$. (For example, the meson exchange current corrections 
in table 7.12 of ref.\cite{ASBH} are on the average $\delta g_{\ell}(p)^{\rm meson}\simeq 0.28$ and 
$\delta g_{\ell}(n)^{\rm meson} \simeq -0.16$, and also the results of ref. \cite{BR} are very similar. 
If eq.(\ref{fh}) would hold only
for the mesonic corrections to $g_{\ell}$, one would obtain a large value $\kappa_A\simeq 0.44$.) However, as is clear
from our derivation, the orbital g-factors which enter in the $\kappa-g_{\ell}$ relation are the {\em total} ones,
including the configuration mixing effects \cite{SHIM} (except for the conventional RPA-type effects 
like first order configuration mixing), besides meson exchange currents.}.  

\section{Connection to the photon scattering amplitude}

In this section we wish to discuss the physical reason which underlies the $\kappa-g_{\ell}$ relation.
In the original discussions on this relation \cite{FUI}-\cite{FIH}, the enhancement factor $\kappa$ was calculated in
perturbation theory directly from the commutator of eq.(\ref{sum2}), and the
orbital g-factor was calculated from the current according to eq.(\ref{gldef}) in the same order of perturbation
theory. In this approach,
the $\kappa-g_{\ell}$ relation appears a bit fortuitously. On the other hand, following the lines
discussed in sect. 4, 
it is clear that the low energy theorem (\ref{let}) directly leads to the $\kappa-g_{\ell}$ relation: The
polarization, which appears on the l.h.s. of this relation, contains the part $\Pi_A$, which is 
related to the p-h excitations as illustrated in Fig. 3, and in the limit of $\omega\rightarrow 0$ 
followed by $|{\bold q}|\rightarrow 0$ these p-h excitations can take place only at the Fermi surface.
It is then immediately clear from Fig. 3 that this process is proportional to the matrix element of the
effective vertex $\Gamma^{(\omega)}$ (the black square in Fig. 3) at the Fermi surface, which directly gives the
quasiparticle current and the orbital g-factors via the definitions (\ref{curr}) and (\ref{gldef}). 
On the other hand, the r.h.s. of (\ref{let}) is related to $\kappa$, which establishes the 
desired relation. 

From our discussions in subsect. 4.2 it is clear that the basic relation (\ref{let}) follows from gauge 
invariance. Actually it can be viewed as a consequence of current conservation applied to the nuclear Compton
scattering amplitude: Eq.(\ref{new0}) shows that the Compton amplitude naturally splits into the 
Thomson amplitude, which follows from the first term in (\ref{new0}), and the ``intrinsic'' scattering amplitude, which 
follows from the Hamiltonian (\ref{hel}) by using the effective charges (\ref{effc}). 
This ``intrinsic'' scattering amplitude $T_{\rm int}^{\mu \nu}$ consists of two pieces,
\begin{eqnarray}
T_{\rm int}^{\mu \nu} = \Pi^{\mu \nu} + M_{\rm int}^{\mu \nu}\,,  \label{scat}
\end{eqnarray}
where $\Pi^{\mu \nu}$ is the current-current correlation function defined in (\ref{pol}) and is often called
the ``resonance amplitude'' in this connection \cite{SCHU2}, and $M_{\rm int}^{\mu \nu}$
is the ``seagull'' amplitude described by the double commutator term in 
(\ref{comm}). Some examples for $M_{\rm int}^{\mu \nu}$ are shown
\footnote{The current operators which 
define the correlation function $\Pi^{\mu \nu}$ include the 2-body exchange currents as specified by the single 
commutator term in $\delta V$ of eq. (\ref{commv}). The piece $M_{\rm int}^{\mu \nu}$ therefore contains all processes 
which cannot be expressed as the product of two current operators.} in Fig. 6.

\begin{figure}[h]
\begin{center}
\epsfig{file=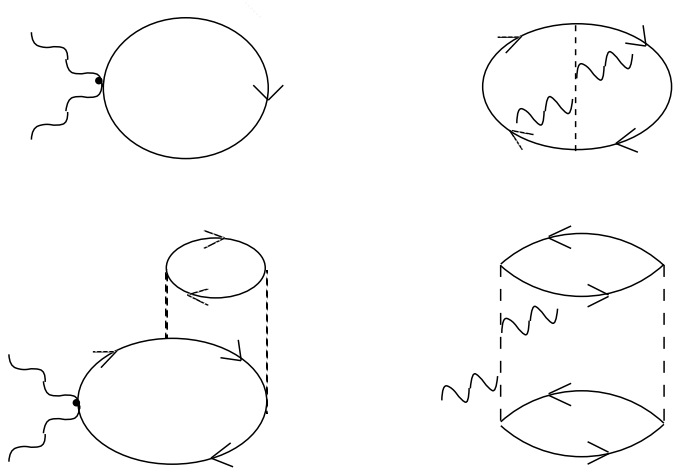,angle=0,width=9cm}
\caption{Some selected examples for diagrams contributing to the seagull amplitude $M_{\rm int}$ of eq.(\ref{scat}).
The seagull-type vertices in the first and third diagrams come from the terms $\propto {\bold A}^2$ in the 
non-interacting part of the Hamiltonian (\ref{hel}), while the vertices in the second and fourth diagrams come from 
the interaction part $\delta V$.}
\end{center}
\end{figure}

Since we have shown that the underlying
Hamiltonian (\ref{hel}) is invariant with respect to local gauge transformations generated by the effective
charges, the amplitude (\ref{scat}) must satisfy current conservation,
in particular
\begin{eqnarray}
q_{\mu} q_{\nu} T_{\rm int}^{\mu \nu}({\bold q}, \omega) = 0\,.  \label{cons}
\end{eqnarray}
Since $\omega$ and ${\bold q}$ in this relation can be treated as independent quantities, 
the limit $\omega\rightarrow 0$ first followed by ${\bold q}\rightarrow 0$ leads to
\begin{eqnarray}
T_{\rm int}^{i j}({\bold q}=0, \omega=0) =  \Pi^{i j}({\bold q}=0, \omega=0) + M_{\rm int}^{i j}({\bold q}=0, \omega=0)
= 0\,. \nonumber \\
\label{effe} 
\end{eqnarray}
Since $M_{\rm int}^{i j}({\bold q}=0, \omega=0)=\langle 0|[[H,D^i],D^j]|0 \rangle$ follows from (\ref{comm}), 
we immediately obtain low energy theorem (\ref{let}). The underlying physical reason for the
$\kappa-g_{\ell}$ relation can therefore be identified as the condition of current conservation for the
nuclear Compton scattering amplitude.

In Appendix E we discuss further connections between the low energy Compton scattering amplitude and
the orbital g-factors, making contact to the analysis in ref.\cite{SCHU2} of photon scattering data.
There we also comment on the sum rule derived a long time ago
by Gerasimov \cite{GER}. It is well known that this relation seemed to
indicate that the (unretarded) E1 sum rule has the same value as the integral over
the total photoabsorption cross section including retardation effects and all higher order
multipoles (see Appendix E). However, its validity has been questioned in a series of papers 
\cite{YAZ}-\cite{ARD}. In particular, in ref. \cite{ARD} it has been shown that the
assumption concerning the high energy behavior of the amplitude is actually not
valid due to the presence of the anomalous magnetic moment term.    

\section{Summary and conclusions}
In this work we used the Landau-Migdal theory to discuss the
orbital angular momentum g-factor of a quasiparticle and the E1 sum rule for isospin
asymmetric nuclear matter. The relations obtained for the orbital g-factors are
in principle exact and hold also in relativistic field theory. 
For the E1 sum rule, we had to restrict ourselves to a nonrelativistic framework because
of the problems arising from the center of mass motion. We discussed in detail the
form of the Hamiltonian which is invariant with respect to local gauge transformations
generated by the effective E1 charges. 

We have split the strength function into two parts, where one comes
from the p-h cuts including the effects of the higher excited states via their real parts, and the other 
comes from cuts at higher excitation energies. We have shown
generally that the former part is related to the orbital g-factors, while the latter part
has no relation to them. The former part has a close relation to the collective excitations of the system, 
i.e., the zero sound modes in infinite systems and the giant resonances in finite nuclei.
We have discussed the importance of the $\kappa-g_{\ell}$ relation, which effectively
separates the observable part of the LWL sum rule, which is related to the strength function
in the low energy region, from the rest.   
Our discussions, which do not rely on perturbation theory, can serve to put many 
previous investigations on the $\kappa-g_{\ell}$ relation on a theoretically firm basis.

Concerning possible extensions, we would like to remark the following points: First, 
the methods used here to relate $\kappa$ to $g_{\ell}$ refer to infinite nuclear matter, 
and it would be interesting
to investigate to what extent they can be applied also to finite nuclei. Second,
as we mentioned in the Introduction, very interesting attempts are now being made to extend the 
range of applicability of the Landau-Migdal theory \cite{SPETH} to give a more general description 
of nuclear collective vibrations. The basic idea is to generalize the definition of the
quantity $A$, which appears in the equation of the vertex (\ref{gfu}) etc., so as to include also
more complicated configurations. It would be
very interesting to see whether the results derived in this paper can be extended according
to these lines.             

\vspace{1cm}

{\sc Acknowledgements}\\
We wish to thank Profs. Y. Horikawa, M. Ichimura, H. Sagawa, M. Schumacher, T. Suzuki (Fukui Univ.),
T. Suzuki (Nihon Univ.) and K. Yazaki for helpful discussions on Giant Resonances and sum rules.
This work was supported by the Grant in Aid for Scientific Research of the Japanese Ministry of
Education, Culture, Sports, Science and Technology, Project No. C2-13640298.

\newpage
\appendix
{\LARGE Appendices}
\section{Ward identities and the relativistic Landau relations for $N\neq Z$}
\setcounter{equation}{0}
In this Appendix we derive some relations used in sect.3, which are based on gauge and Lorentz invariance.
For more detailed discussions for the case of isospin symmetric nuclear matter, we refer to 
refs.\cite{BEN,BC}.

\subsection{Ward identities and currents}
If $Q_p$ and $Q_n$ are the bare electric charges for protons and neutrons, the requirement of
local gauge invariance leads to the Ward-Takahashi identity between the electromagnetic vertex 
$\Gamma_{\alpha}$ and the propagator $S_{\alpha}$ for protons ($\alpha=p$) and neutrons ($\alpha=n$):
\begin{eqnarray}
q_{\mu} \Gamma_{\alpha}^{\mu}(k',k) = Q_{\alpha} \left(S_{\alpha}^{-1}(k') - S_{\alpha}^{-1}(k)\right)\,. 
\label{wth}
\end{eqnarray}
By setting $q^0=0$ first and then letting ${\bold q}\rightarrow 0$, one obtains the Ward identity
for the vertex in the ``static limit'' \cite{MIGB}:
\begin{eqnarray}
{\bf \Gamma}_{\alpha}^{({\rm st})}(k,k) = - Q_{\alpha} {\bold \nabla}_k\,S_{\alpha}^{-1}(k)\,. \label{w}
\end{eqnarray}
If the propagator has a quasiparticle pole $k_0=\epsilon_{\alpha}(k)$ with residue $Z_{\alpha}(k)$, 
we can expand $S_{\alpha}^{-1}(k) = Z_{\alpha}(k)^{-1} \left(k_0 - \epsilon_{\alpha}(k)\right)
+ {\cal O}\left[\left(k_0-\epsilon_{\alpha}(k)\right)^2\right]$ to obtain
the current in the static limit on the quasiparticle energy shell ($k_0 \rightarrow \epsilon_{\alpha}(k)$)
as follows\footnote{In this Appendix, we
will denote the current for the charges $Q_{\alpha}$ as $J$, which corresponds to $j$ of sect. 3
for the physical charges, and to ${\hat j}$ of sect. 4 for the effective charges.}: 
\begin{eqnarray}
{\bold J}_{\alpha}^{({\rm st})}(k) \equiv Z_{\alpha}(k)\, {\bf \Gamma}_{\alpha}^{({\rm st})}(k,k)
= Q_{\alpha} \, {\bold \nabla}_k \,\epsilon_{\alpha}(k) \label{cst}
\end{eqnarray}
On the Fermi surface, this gives the static current as ${\bold J}_{\alpha}^{({\rm st})}(k)={\hat {\bf k}}\,
Q_{\alpha} \, v_F(\alpha)$, where $v_F(\alpha)$ is the Fermi velocity for protons or neutrons.

If we take the static limit ($q_0 \rightarrow 0$ first, then $|{\bold q}|\rightarrow 0$) in eq.(\ref{curr1}) for
$|{\bold k}|$ on the Fermi surface, and use the form (\ref{fermis}) for the p-h propagator,  
we can use the above form of the current in the static limit on the l.h.s. and the r.h.s. under the integral. 
In this kinematics the interaction $t^{(\omega)}$ depends only on the angle between the directions
of ${\bold k}$ and ${\bold \ell}$. One then obtains the current in the ``$\omega$ limit'' as
\begin{eqnarray}
{\bold J}^{(\omega)}_{\alpha}(k) = {\hat {\bold k}} \, Q_{\alpha}\, v_F(\alpha) + 2 \sum_{\beta=p,n}
\int \frac{{\rm d}\Omega_{\ell}}{(2\pi)^3}\, f_{\alpha \beta}({\hat {\bold k}} \cdot {\hat {\bold \ell}}) \,
p_F^2(\beta) \, Q_{\beta} \, {\hat {\bold \ell}}\,,  \label{eq}
\end{eqnarray}
where $f$ is the spin independent part of $t^{(\omega)}$. The angular integrations are trivial, and if one uses
the definition of the dimensionless interaction (\ref{deff}) one obtains the currents (\ref{jp}) and (\ref{jn}) for the
physical charges, or (\ref{jp1}) and (\ref{jn1}) for the effective charges, where $F_1$
are the coefficient of the ${\ell}=1$ Legendre polynomial in the expansion of the spin independent part
of the interaction ${\cal F}$. 
 
\subsection{Lorentz (Galilei) invariance for $N\neq Z$}
\setcounter{equation}{0}
The requirements of Lorentz invariance have been discussed in ref.\cite{BC}, and are easily generalized
to the case $N\neq Z$ as follows:

If we observe matter from a system ${\tilde S}$ which moves with velocity $-{\bold v}$ relative to the system $S$ where
matter is at rest, the Fermi distribution function of neutrons ($\alpha=n$) or protons ($\alpha=p$) 
for fixed momentum ${\bold \ell}$ will change by an amount
${\tilde n}_{\ell}(\alpha) -n_{\ell}(\alpha)$, where ${\tilde n}$ and $n$ are the Fermi distributions in 
${\tilde S}$ and $S$. According to Landau's basic hypothesis \cite{LAND},
this corresponds to a change of the quasiparticle energy (for fixed momentum ${\bold k}$) by an 
amount\footnote{Since the quasiparticle interaction 
$f_{\alpha \beta} (k', \ell)$ used in (\ref{hyp}) refers to the system $S$, the relation (\ref{hyp}) actually
holds only up to the first order in ${\bold v}$, which is sufficient for our purpose. 
Concerning the Lorentz transformation of the quasiparticle interaction, see ref.\cite{BC}.}  
\begin{eqnarray}
{\tilde \epsilon}_{k'}(\alpha) - \epsilon_{k'}(\alpha) = 2 \int \frac{{\rm d}^3 p}{(2\pi)^3}
\sum_{\beta=p,n} f_{\alpha \beta} (k', \ell) \left({\tilde n}_{\ell}(\beta) -n_{\ell}(\beta)\right)\,,  
\label{hyp}
\end{eqnarray}
where $f_{\alpha \beta}$ is the spin independent part of the quasiparticle interaction 
$t_{\alpha \beta}^{(\omega)}$, as in (\ref{eq}). On the other hand, the l.h.s. of
(\ref{hyp}) must agree with the result obtained from a Lorentz transformation:
\begin{eqnarray}
{\tilde \epsilon}_{k'}(\alpha) = \gamma \left(\epsilon_{k}(\alpha) + {\bold k}\cdot {\bold v}\right)\,, \label{l1}
\end{eqnarray}
where ${\displaystyle \gamma=(1-{\bold v}^2)^{\frac{1}{2}}}$, 
and ${\bold k}'$ and ${\bold k}$ are related by the Lorentz transformation
\begin{eqnarray}
{\bold k}' = {\bold k} + \epsilon_k(\alpha) {\bold v} \gamma - {\hat {\bold v}} \left({\hat {\bold v}} \cdot
{\bold k} \right) \left(1-\gamma\right)\,.  \label{l2}
\end{eqnarray} 
For the r.h.s. of (\ref{hyp}), we use the fact that the Fermi distribution is Lorentz invariant:
\begin{eqnarray}
{\tilde n}_{\ell'}(\alpha) = n_{\ell}(\alpha) \label{l3}  
\end{eqnarray}
where ${\bold \ell}'$ and ${\bold \ell}$ are related by a Lorentz transformation analogous to (\ref{l2}).

To first order in ${\bold v}$ one easily obtains from (\ref{l1}), (\ref{l2}) and (\ref{l3}):
\begin{eqnarray}
{\tilde \epsilon}_{k'}(\alpha) - \epsilon_{k'}(\alpha) &=& {\bold k}\cdot {\bold v} - \epsilon_k(\alpha)
{\bold v}\cdot {\bold v}_k(\alpha)  \label{t1p} \\
{\tilde n}_{\ell}(\alpha) -n_{\ell}(\alpha) &=& \epsilon_{\ell}(\alpha) {\bold v}\cdot {\hat {\bold \ell}}\,\, 
\delta(|{\bold \ell}|-p_F(\alpha))\,,
\label{t2p}
\end{eqnarray}
where ${\bold v}_k(\alpha)= {\bold \nabla}_k \epsilon_k(\alpha)$ is the quasiparticle velocity.
Inserting the relations (\ref{t1p}) and (\ref{t2p}) into (\ref{hyp}) and setting $|{\bold k}|=p_F(\alpha)$, 
we obtain 
\begin{eqnarray}
v_F(p) + \frac{v_F}{6} \left[\left(\frac{p_F(p)}{p_F}\right)^2 F_1(pp) + \frac{\mu_n}{\mu_p}
\left(\frac{p_F(n)}{p_F}\right)^2 F_1(pn) \right] &=& \frac{p_F(p)}{\mu_p} \nonumber \\  
v_F(n) + \frac{v_F}{6} \left[\left(\frac{p_F(n)}{p_F}\right)^2 F_1(nn) + \frac{\mu_p}{\mu_n}
\left(\frac{p_F(p)}{p_F}\right)^2 F_1(pn) \right] &=& \frac{p_F(n)}{\mu_n}  \nonumber \\
\label{end} 
\end{eqnarray}
where $\mu_{\alpha}=\epsilon_F(\alpha)$ are the chemical potentials (Fermi energies) of protons and neutrons.
These relations were used to derive eqs.(\ref{gp}), (\ref{gn}) and (\ref{hat}) in the main text.

\section{Local gauge invariance of the Hamiltonian (\ref{comm})}
\setcounter{equation}{0}
Here\footnote{To simplify the notations of this Appendix, we will not indicate the dependence of the Hamiltonian
and the dipole operator on the charges $Q_p$ and $Q_n$.}  
we show the invariance of the Hamiltonian (\ref{comm}) under local gauge transformations
up to ${\cal O}(e^2)$. For this, we note that the arguments given below eq.(\ref{th1}) in the main text show that
for any operator ${\cal B}$, which commutes with the z-component of the total isospin operator ($[T_z,{\cal B}]=0$),
the gauge transformations like (\ref{th0}) or (\ref{th1}) can be expressed by the dipole operator. That is, if
$[T_a,{\cal B}]=0$, the gauge transformation of ${\cal B}$ can be expressed as follows:
\begin{eqnarray}
{\cal B} \rightarrow {\rm exp}\left[-i {\bold c}\cdot {\bold D}\right] \, {\cal B}\, 
{\rm exp}\left[i {\bold c}\cdot {\bold D}\right]
= {\cal B} -i c^i [D^i,{\cal B}] - \frac{1}{2} c^i c^j [D^i,[D^j,{\cal B}]] + \dots
\nonumber \\ \label{o}
\end{eqnarray}
We have to apply this transformation, together with 
\begin{eqnarray}
{\bold A} \rightarrow {\bold A}+{\bold c} \label{ga}
\end{eqnarray}
to the Hamiltonian (\ref{comm}). First, since $[T_z,H]=0$, we can use eq.(\ref{o}) for ${\cal B}=H$, as we have done
already in (\ref{haus}) of the main text. Second, also for the single commutator term on the r.h.s. of (4.12) we have
$[T_z,[D^i,H]]=0$, as can be seen easily from the Jacobi identity for double commutators. Third, the last term
in (\ref{comm}) is already of ${\cal O}(e^2)$, and therefore it is affected only by the gauge transformation
of ${\bold A}$, eq.(\ref{ga}). We therefore obtain for the gauge transformation of the Hamiltonian (\ref{comm}):   
\begin{eqnarray}
{\cal H} &\rightarrow& {\rm exp}\left[-i {\bold c}\cdot {\bold D}\right] \left(H + i \left(A^j+c^j\right) [D^j,H] \right) 
{\rm exp}\left[i {\bold c}\cdot {\bold D}\right]  \nonumber \\
&-& \frac{1}{2} \left(A^i A^j+2 A^i c^j + c^i c^j\right) [D^i,[D^j,H]] \label{trh}
\end{eqnarray}
Using then (\ref{o}) for ${\cal B}=H$ and ${\cal B}= i [D^i,H]$, it is easy to see that we simply get back 
the three terms on the r.h.s. of (\ref{comm}), which shows the gauge invariance of ${\cal H}$.

\section{Time-ordered diagrams}
\setcounter{equation}{0}
In this Appendix, we first explain the three points stated in sect. 5 concerning the relation between 
$\Pi_A$, which was originally defined in the language of Feynman diagrams in sect.2, and the time-ordered
Bethe-Goldstone diagrams. Although the arguments presented below can be generalized by using dispersion representations
for the vertex, the interaction and the self energy, we prefer to discuss representative examples 
in order to be definite. The arguments can be applied immediately also to other cases. We then will derive the 
relation (\ref{disa}).

\subsection{2p-2h states induced by vertex corrections}
The third diagram in Fig. 4 arises from the contribution to the quasiparticle vertex $\Gamma^{(\omega)}$ which
is shown in Fig. 7. By performing the frequency integrals for the two loops, one obtains the
energy denominators
\begin{eqnarray}
\left[ \left(k_0+\frac{\omega}{2}-E_1\right)\left(k_0-\frac{\omega}{2}-E_2\right)\right]^{-1}\,,  \label{d1} 
\end{eqnarray}
where the energies $E_1$ and $E_2$ are indicated in Fig.7. 

\begin{figure}[h]
\begin{center}
\epsfig{file=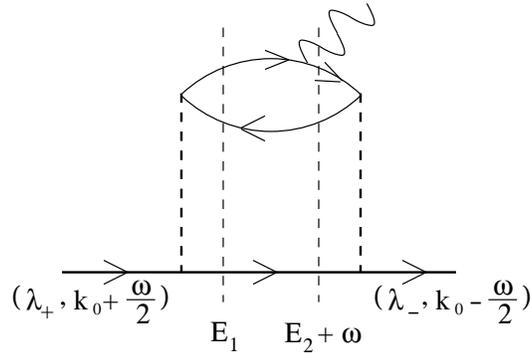,angle=0,width=7cm}
\caption{A contribution to the vertex $\Gamma^{(\omega)}$, which is associated with the factor (\ref{d1}).}
\end{center}
\end{figure}

If this vertex appears in a Feynman
diagram together with the p-h propagator $A$ of eq.(\ref{al}),
the effect of $A$ is to replace the energy denominator (\ref{d1}) by 
\begin{eqnarray}  
\left[ \left(\epsilon_{\lambda_+}- E_1\right)\left(\epsilon_{\lambda_-}-E_2\right)\right]^{-1}
\label{d2}
\end{eqnarray}
This is the ``high energy denominator'' of $\Pi_A$. 
On the other hand, the high energy denominator in the time-ordered diagram (third diagram of Fig. 4) is
\begin{eqnarray}
\left[ \left(\omega+\epsilon_{\lambda_-}-E_1\right)\left(\epsilon_{\lambda_-}-E_2\right)\right]^{-1}  \label{d3}
\end{eqnarray}
If we replace  $\omega\rightarrow E^{\rm ph}\equiv \epsilon_{\lambda_+}- \epsilon_{\lambda_-}$ in (\ref{d3}),
the expression agrees with (\ref{d2}). In other words, if we replace in the
high energy denominator of the time-ordered graph $\omega$ by the excitation energy
of the p-h pair which enters (or leaves) the vertex, we obtain the contribution of this graph to $\Pi_A$.
This is the content of point 1. discussed in sect. 5. 

\subsection{2p-2h states induced by the interaction}
The fifth diagram in Fig. 4 arises from the contribution to the quasiparticle interaction $T^{(\omega)}$ which
is shown in Fig. 8. By performing the frequency integral for the loop, one obtains the
energy denominator
\begin{eqnarray}
\left[k_0-\ell_0 - E \right]^{-1}\,,  \label{d1a} 
\end{eqnarray}
where the energy $E$ is indicated in Fig. 8. 

\begin{figure}[h]
\begin{center}
\epsfig{file=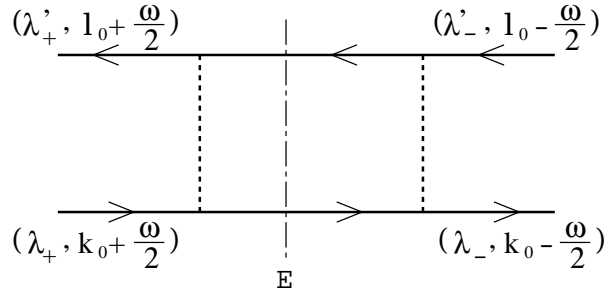,angle=0,width=8cm}
\caption{A contribution to the interaction $T^{(\omega)}$, which is associated with the factor (\ref{d1a}).}
\end{center}
\end{figure}

If this interaction appears in a Feynman
diagram together with two p-h propagators $A$, which correspond to the incoming and
outgoing p-h pairs, the effect of $A$ (see eq.(\ref{al})) is to replace the energy denominator (\ref{d1}) by 
\begin{eqnarray}  
\left[\frac{1}{2} \left(\epsilon_{\lambda_+}+\epsilon_{\lambda_-} - \epsilon_{\lambda'_+} - \epsilon_{\lambda'_-} \right)
-E \right]^{-1} \,.\label{d2a}
\end{eqnarray}
This is the ``high energy denominator'' of $\Pi_A$. 
On the other hand, the high energy denominator in the time-ordered diagram (fifth diagram of Fig. 4) is
\begin{eqnarray}
\left[ \omega + \epsilon_{\lambda_-} - \epsilon_{\lambda'_+} - E \right]^{-1}\,.  \label{d3a}
\end{eqnarray}
If we replace $\omega\rightarrow \frac{1}{2} \left(E_i^{\rm ph}+ E_f^{\rm ph}\right) 
\equiv \frac{1}{2} \left(\epsilon_{\lambda_+} - \epsilon_{\lambda_-} + \epsilon_{\lambda'_+} - \epsilon_{\lambda'_-}\right)$ 
in (\ref{d3a}), the expression agrees with (\ref{d2a}). In other words, if we replace in the
high energy denominator of the time-ordered graph $\omega$ by the average of the excitation energies
of the p-h pair which enters and leaves the interaction, we obtain the contribution of this graph to $\Pi_A$.
This is the content of point 2. discussed in sect. 5.

\subsection{2p-2h states induced by the self energy}
The p-h propagator $A$ defined in sect.2 includes by definition the effects of the self energies on the
quasiparticle energies and the residues. For example, if we denote by $\gamma$ the non-interacting electromagnetic
vertex, the contribution $\left(i \gamma A \gamma\right)$ to $\Pi_A$ in eq.(2.10) includes the following factor 
which is associated with the diagram shown in Fig. 9a:
\begin{eqnarray}
\frac{\Sigma(E_{\lambda_+})}{\left(\omega-E_{\lambda_+}+E_{\lambda_-}\right)^2} + 
\frac{\partial \Sigma/ \partial k_0 (E_{\lambda_+})}{\omega-E_{\lambda_+}+E_{\lambda_-}}\,.  \label{d0b}
\end{eqnarray}
Here the energies $E_{\lambda}$ do not include the effect of the dispersive self energy, which 
is shown in Fig. 9b. After performing the frequency integrals for the two loops in the Feynman diagram
for $\Sigma$, one obtains the energy denominator
\begin{eqnarray}
\left(k_0 + \frac{\omega}{2} - E \right)^{-1}\,,  \label{d1b} 
\end{eqnarray}
where the energy $E$ is indicated in Fig. 9b.

\begin{figure}[h]
\begin{center}
\epsfig{file=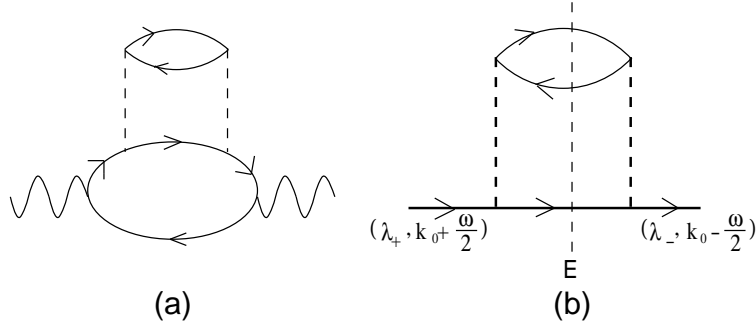,angle=0,width=10cm}
\caption{(a) A contribution of dispersive self energies to the polarization. (b) A dispersive self energy
part, which is associated with the factor (\ref{d1b}).}
\end{center}
\end{figure}

Using this in eq.(\ref{d0b}), we see that the high energy denominators in $\Pi_A$ have the form
\begin{eqnarray}
\left[ \left(\omega-E_{\lambda_+}+E_{\lambda_-}\right)^2 \left(E_{\lambda_+} - E\right) \right]^{-1}
- \left[\left(\omega-E_{\lambda_+}+E_{\lambda_-}\right) \left(E_{\lambda_+} - E\right)^2 \right]^{-1}\,.
\nonumber \\
\label{d2b}
\end{eqnarray}
On the other hand, the high energy denominators of the time-order diagram for the polarization are
\begin{eqnarray}
\left[ \left(\omega-E_{\lambda_+}+E_{\lambda_-}\right)^2 \left(\omega + E_{\lambda_-} - E\right) \right]^{-1}
\label{d3b}
\end{eqnarray}
Expanding the high energy denominator (second factor in (\ref{d3b})) around $\omega=E_{\lambda_+}-E_{\lambda_-}$
and taking only the first two terms of this expansion, we obtain an expression which agrees with (\ref{d2b}).
In other words, if we replace the high energy denominator of the time-ordered graph by its first two
terms of an expansion around the excitation energy of the p-h pair under consideration, we obtain the contribution 
of this graph to $\Pi_A$. This is the content of point 3. discussed in sect. 5.

\subsection{Proof of relation (\ref{dis})}

For each time-order graph which contributes to the polarization, there is another one which 
is obtained from the original graph by crossing the photon lines. That is, each graph $\Pi^{(i)}$, which is
characterized by a particular topology of the internal propagators and two-body interactions, can be split into two parts, 
$\Pi^{(i)}=\Pi^{(i)}_+ + \Pi^{(i)}_-$, where $\Pi^{(i)}_+$ does not involve crossed photons (``forward term''), and      
$\Pi^{(i)}_-$ involves crossed photons (``backward term''). By definition, the intermediate states in $\Pi^{(i)}_+$
have either no photons (energy denominators of the form $\left(\omega-E_n + i \epsilon\right)^{-1}$), or one photon (energy
denominators are independent of $\omega$). On the other hand, the intermediate states in $\Pi^{(i)}_-$
have either two photons (energy denominators of the form $\left(\omega+E_n - i \epsilon\right)^{-1}$), or one photon (energy
denominators are independent of $\omega$). It follows that $\Pi^{(i)}_+(\omega)$ is analytic in the upper $\omega$ plane
\cite{TAKA}, and has a non-vanishing imaginary part only for real positive $\omega$. From these properties it follows that
$\Pi^{(i)}_+(\omega)$ satisfies a dispersion relation
\begin{eqnarray}
\Pi^{(i)}_+(\omega) = \frac{-1}{\pi} \int_0^{\infty} \frac{{\rm d}\omega'}{\omega-\omega' + i \epsilon}\,\,
{\rm Im}\,\, \Pi^{(i)}_+(\omega')\,.  \label{r1}
\end{eqnarray}     
Since $\Pi^{(i)}_+(\omega)$ and $\Pi^{(i)}_-(\omega)$ differ only by the external photon lines,
they give the same contribution at $\omega=0$, that is, $\Pi^{(i)}_+(\omega=0) = \Pi^{(i)}_-(\omega=0)$. 
We therefore obtain from eq.(\ref{r1}) in the limit $\omega=0$
\begin{eqnarray}
\lefteqn{\int_0^{\infty} \frac{{\rm d}\omega}{\omega}\,\, {\rm Im}\, \Pi^{(i)}(\omega) =
\int_0^{\infty} \frac{{\rm d}\omega}{\omega}\,\, {\rm Im}\, \Pi^{(i)}_+(\omega)} \nonumber \\
& & = \pi\,\Pi^{(i)}_+(\omega=0) = \frac{\pi}{2}\, \Pi^{(i)}(\omega=0)\,, \label{r2}
\end{eqnarray}
which is the content of eq.(\ref{dis}).

\section{Connection between $\Pi_A$ and the collective excitations}
\setcounter{equation}{0}
In this Appendix we derive the expression for the polarization $\Pi_A$ 
in terms of the solutions to the Landau equation, referring for simplicity 
to symmetric nuclear matter (N=Z) and low energy and momentum transfer, 
where $\Pi_A$ of eq.(\ref{Piaf}) depends only on the ratio $s=\omega/(|{\bold q}| v_F)$.
For the case N=Z, the effective charges (\ref{effc}) become $\pm \frac{1}{2} e_p$ for protons
and neutrons, and therefore the currents in the polarization (\ref{Piaf}) are the
isovector (IV) currents. 

If we approximate the p-h propagator in (\ref{Piaf}) by (\ref{fermis}), the quantity
\begin{eqnarray}
{\bold \phi}({\bold k},s) &\equiv& \frac{{\bold q}\cdot {\hat {\bold k}} \, \delta(k-p_F)}
{\omega-{\bold q}\cdot {\bold v}_k + i \gamma}\,{\bold j}_{\rm IV}(k_+, k_-)  \equiv 
\delta(|{\bold k}|-p_F)\, {\bold \phi}({\hat {\bold k}},s)  \nonumber \\  \label{phwf}
\end{eqnarray}
satisfies the equation
\begin{eqnarray}
\left(s - \cos \Theta_k \right) {\bold \phi}({\hat {\bold k}},s) = \frac{\cos \Theta_k}{v_F}\, 
{\bold j}_{\rm IV}^{(\omega)}
({\hat {\bold k}}) + \frac{\cos \Theta_k}{4 \pi}  \int {\rm d} \Omega_{\ell}\,
F'({\hat {\bold k}}\cdot {\hat {\bold \ell}})\,{\bold \phi}({\hat {\bold \ell}},s)\,, \nonumber \\
\label{qp}
\end{eqnarray}
where $\cos \Theta_k \equiv {\hat {\bold q}}\cdot {\hat {\bold k}}$, $F'({\hat {\bold k}}\cdot {\hat {\bold \ell}})$
is the spin independent part $\propto ({\bold \tau}_1 \cdot {\bold \tau}_2)$ of the dimensionless Landau-Migdal
interaction (\ref{deff}), and 
${\bold j}^{(\omega)}_{\rm IV}({\hat {\bold k}})$ is the isovector part of the quasiparticle current in the $\omega$-limit
at the Fermi surface, i.e., from eqs. (\ref{gp}), ({\ref{gn}): 
\begin{eqnarray}
{\bold j}^{(\omega)}_{\rm IV}({\hat {\bold k}}) \equiv {\hat {\bold k}} \, j^{(\omega)}_{\rm IV}
= {\hat {\bold k}}\, \frac{1}{2} \, \left(\frac{p_F}{M}-\frac{v_F}{3}\left(F_1-F_1'\right)\right).  \label{iv}
\end{eqnarray} 
For small $|{\bold q}|$ it is sufficient to seek the solution to (\ref{qp}) in the form 
${\bold \phi}({\hat {\bold k}},s) \equiv {\hat {\bold k}} \, \phi ({\hat {\bold k}},s)$, where the scalar function
$\phi ({\hat {\bold k}},s)$ satisfies an equation similar to (\ref{qp}), but with the interaction 
retarded by one unit of angular momentum:
${\cal F}({\hat {\bold k}}\cdot {\hat {\bold \ell}}) \rightarrow {\cal F}({\hat {\bold k}}\cdot {\hat {\bold \ell}})
\, {\hat {\bold k}}\cdot {\hat {\bold \ell}}$. We can write the equation for $\phi ({\hat {\bold k}},s)$ in the
form 
\begin{eqnarray}
\int {\rm d} \Omega_{\ell}\,K({\hat {\bold k}},{\hat {\bold \ell}})\, \phi({\hat {\bold \ell}},s) + \frac{\cos \Theta_k}{v_F} 
j_{\rm IV}^{(\omega)} = s\, \phi({\hat {\bold k}},s)  \label{inh}
\end{eqnarray}
with the kernel
\begin{eqnarray}
K({\hat {\bold k}},{\hat {\bold \ell}}) \equiv \cos \Theta_k\, \delta(\Omega_k-\Omega_{\ell}) + 
\frac{\cos \Theta_k}{4 \pi} 
F'({\hat {\bold k}}\cdot {\hat {\bold \ell}})\,{\hat {\bold k}}\cdot {\hat {\bold \ell}}.
\end{eqnarray}
One can express the solution to the inhomogeneous equation (\ref{inh}) by the solutions to the
homogenoues equations
\begin{eqnarray}
\int {\rm d} \Omega_{\ell}\,K({\hat {\bold k}},{\hat {\bold \ell}})\, \Phi({\hat {\bold \ell}},s)  &=& s\,
\Phi({\hat {\bold k}},s)  \label{hr} \\
\int {\rm d} \Omega_{k}\, \overline{\Phi}({\hat {\bold k}},s)\,K({\hat {\bold k}},{\hat {\bold \ell}})  &=& s\,
{\overline \Phi}({\hat {\bold \ell}},s)\,,  \label{hl}  
\end{eqnarray}
where we must distinguish between the right and left eigenfunctions because of the nonhermiticity of the
kernel $K$. The left eigenfunction is related to the ordinary h.c. of $\Phi$ by 
${\displaystyle \overline{\Phi}({\hat {\bold k}},s) = \Phi^{\dagger}({\hat {\bold k}},s) \frac{N_s}{\cos \Theta_k}}$,
where $N_s$ is chosen to satisfy the normalization (\ref{norm}). 

For $|s|<1$ there exist ``free'' 
solutions to eq.(\ref{hr}) and (\ref{hl}), namely 
${\displaystyle \Phi({\hat {\bold k}},s) = {\overline \Phi}({\hat {\bold k}},s) = \frac{1}{\sqrt{2\pi}} 
\delta(s- \cos \Theta_k)}$, and in this case the homogeneous equations can be solved iteratively
for any value $|s|<1$, i.e., there is a continuum of solutions. For $|s|>1$, on the other hand, 
no free solutions exist, and the eigenvalue $s$ becomes discrete. These are the zero sound solutions,
which have been investigated in detail in many works \cite{TBA,YH,NOZ}. 

We can choose the right and left eigenfunctions so as to form a
complete orthonormal set:
\begin{eqnarray}
\lefteqn{\int {\rm d}\Omega_k\, {\overline \Phi}({\hat {\bold k}},s') \, \Phi({\hat {\bold k}},s) = \delta_{s',s}}
\label{norm} \\ 
& & {\sum}_s \Phi({\hat {\bold k}},s)\,{\overline \Phi}({\hat {\bold \ell}},s) = 
\delta\left(\Omega_k-\Omega_{\ell}\right)\,,
\label{comp}
\end{eqnarray}
where $\delta_{s',s}$ stands for a delta function in the case of continous eigenvalues, and the Kronecker delta
symbol in the case of discrete ones. The symbol $\sum_s$ refers to an integral over continuous eigenvalues,
and a sum over discrete ones.
The solution to the inhomogeneous equation (\ref{inh}) can then be expressed as
follows:
\begin{eqnarray}
\phi({\hat {\bold k}},s) = \frac{j_{\rm IV}^{(\omega)}}{v_F}\, {\sum}_{s'}\, \int {\rm d} \Omega_{\ell}\, 
\frac{\Phi({\hat {\bold k}},s') {\overline \Phi}({\hat {\bold \ell}},s')}{s-s'+i\gamma}\,\cos \Theta_{\ell}\,, 
\label{sol}
\end{eqnarray}
where $\cos \Theta_{\ell} = {\hat {\bold q}}\cdot {\hat {\bold \ell}}$.  
Returning to the correlation function (\ref{Piaf}), we obtain in the limit of low $(\omega,{\bold q})$
the form $\Pi_A^{ij}(\omega,{\bold q}) = \delta^{ij} \Pi_A(\omega,{\bold q})$, where $\Pi_A$ is given by 
\begin{eqnarray}
\Pi_A(\omega,{\bold q}) &=& - \frac{V p_F^2}{6 \pi^3} \, j_{\rm IV}^{(\omega)} \,
\int {\rm d}\Omega_k \, \phi({\hat {\bold k}},s) 
= - \frac{A}{4 p_F v_F} \, \left(j_{\rm IV}^{(\omega)}\right)^2 \, \frac{{\sum}_{s'} \, A_{s'}\,B_{s'}} 
{s-s' + i \gamma} \,, \nonumber \\
\label{re1}
\end{eqnarray}
where 
\begin{eqnarray}
A_{s'} &=& \int {\rm d}\Omega_k \, \Phi({\hat {\bold k}},s') \nonumber \\ 
B_{s'} &=& \int {\rm d}\Omega_k \, {\overline \Phi}({\hat {\bold k}},s')\,\cos \Theta_{k}\,=
A_{s'}^*\,N_{s'}\,. 
\label{ab}
\end{eqnarray}
From the homogeneous equations (\ref{hr}) and (\ref{hl}) it is easy to see that by reversing the
direction of the momentum one obtains a solution with the opposite sign of the eigenvalue:
$\Phi(-{\hat {\bold k}},s) = \Phi({\hat {\bold k}},-s)$ and    
${\overline \Phi}(-{\hat {\bold k}},s) = {\overline \Phi}({\hat {\bold k}},-s)$. One can therefore
restrict the summation over the eigenvalues in (\ref{re1}) to positive ones and write
\begin{eqnarray}
\Pi_A(\omega,{\bold q}) = - \frac{A}{4 p_F v_F} \, \left(j_{\rm IV}^{(\omega)}\right)^2 
\,{\sum}_{s'} \, |A_{s'}|^2\, N_{s'} \,\left(\frac{1}{s-s' + i \gamma}- \frac{1}{s+s' -i  \gamma}\right) \,.
\nonumber \\  \label{re2}
\end{eqnarray}

\section{Some relations for the Compton scattering amplitude}
\setcounter{equation}{0}
In this Appendix we discuss relations between the low energy Compton scattering amplitude and the orbital
g-factors (or the enhancement factor $1+\kappa_A$), and make contact to the sum rule derived a long time
ago by Gerasimov \cite{GER}.

\subsection{Relations for the low energy scattering amplitudes}
If we add the Thomson amplitude 
\begin{eqnarray}
M^{ij}_{\rm Th} = \delta^{ij}\,\, \frac{-Z^2\,e_p^2}{MA}  \label{thoms}
\end{eqnarray}
to the intrinsic part (\ref{scat}), we get the total photon scattering amplitude 
\begin{eqnarray}
T ^{i j}({\bold q}, \omega) = \Pi^{i j}({\bold q}, \omega) + M^{i j}({\bold q}, \omega)
\label{total}
\end{eqnarray}
with $M \equiv M_{\rm int} + M_{\rm Th}$. 
Similar to the correlation function $\Pi$, we can split also the seagull amplitude into two parts
as $M = M_A + M_B$. (Examples for 
$M_A$ are shown by the first three diagrams in Fig. 6, while the fourth diagram contributes to $M_B$.)

According to eqs.(\ref{erg}), (\ref{rel}), and (\ref{effe}), the ``A-parts'' of the scattering amplitude (\ref{total}) can be
expressed in terms of the orbital g-factors, or in terms of $\kappa_A$, as follows:
\begin{eqnarray}
\Pi^{i j}_A({\bold q}=0, \omega=0) &=& \delta^{ij} \,e_p^2\, \frac{NZ}{AM}\left(g_{\ell}(p)-g_{\ell}(n)\right)
= e_p^2\,\frac{NZ}{AM} \left(1+\kappa_A\right) \nonumber \\ \label{one}
\\
M^{i j}_A({\bold q}=0, \omega=0) &=& -\Pi_A^{i j}({\bold q}=0, \omega=0)-\frac{Z^2\,e_p^2}{MA} \nonumber \\
&=& - \frac{Z\,e_p^2}{M} g_{\ell}(p) = - \frac{Z\,e_p^2}{M}\left(1 + \frac{N}{A} \kappa_A\right).
\label{two}
\end{eqnarray}
These relations between the low energy nuclear Compton scattering amplitudes and the E1 enhancement factor
have been used extensively in the analysis of experimental data \cite{SCHU2}. One should keep
in mind, however, that they are valid only for the ``A-parts''.  
The ``B-parts'' of $\Pi$ and $M$, which cancel each other in the low energy limit, have no relations
to the orbital g-factors.   

\subsection{Relation to Gerasimov's sum rule} 
In sect. 4 we made use of a spectral representation of the correlation function for fixed ${\bold q}$, 
see eq.(\ref{kl}). Although for a real
photon one has $|{\bold q}|=\omega$, in the energy region where $\omega\, R<<1$ one can
approximate $|{\bold q}|\simeq 0$, leaving $\omega$ finite. This leads to the unretarded
E1 sum rule (\ref{sum1}), (\ref{sum2}). On the other hand, there exists a subtracted dispersion
relation for the full forward scattering amplitude $T(\omega)\equiv T_{ij}({\bold q},\omega)
\,e_i\,e_j$ for a physical photon ($|{\bold q}|=\omega$) 
\cite{GGT}:
\begin{equation}
{\rm Re}\, T(\omega)-T(\omega=0) = - \frac{\omega^2}{2 \pi^2} P \int_0^{\infty} dE\,
\frac{1}{\omega^2-E^2}\, \sigma_{\rm tot}(E).
\label{disp}
\end{equation}
This dispersion relation is often used to extract the total
photoabsorption cross section $\sigma_{\rm tot}$ from the measured elastic cross section
\cite{SCHU1,DAL,SCHU2}.

According to eq.(\ref{total}), the amplitude $T$ consists of the ``resonance amplitude'' 
$\Pi(\omega)\equiv \Pi_{ij}(\omega) \,e_i\,e_j$ and the ``seagull amplitude'' 
$M(\omega)\equiv M_{ij}(\omega) \,e_i\,e_j$. In the energy region beyond the GDR and below the
pion production threshold ($\omega_{\rm GDR}<<\omega<<\omega_{\pi}$),
the energy dependence of the seagull amplitude $M$ (fig.6) is weak: $M(\omega)\simeq M(\omega=0)$.
Therefore one can replace $T \rightarrow \Pi$ on the l.h.s. of (\ref{disp}) to get the relation
\begin{equation}
{\rm Re}\, \Pi(\omega)-\Pi(\omega=0) = - \frac{\omega^2}{2 \pi^2} P \int_0^{\infty} dE\,
\frac{1}{\omega^2-E^2}\, \sigma_{\rm tot}(E)\,, 
\label{disp1}
\end{equation}
which is valid for $\omega_{\rm GDR}<<\omega<<\omega_{\pi}$.
If one makes the further assumption that
in this energy region ($\omega>>\omega_{GDR}$) the resonance amplitude $\Pi(\omega)$ is already small 
because of large energy denominators, one might use (\ref{disp1}) formally also for 
$\omega\rightarrow \infty$ to obtain
\begin{equation}
\Pi(\omega=0) \stackrel{?}{=}  \frac{1}{2 \pi^2} P \int_0^{\infty} dE\, \sigma_{\rm tot}(E)\,,  \label{ger} 
\end{equation}
and comparison with the exact relation (\ref{suma}) would lead to 
\begin{eqnarray}
\int_{0}^{\infty} dE\, S(E) \stackrel{?}{=} \int_{0}^{\infty} dE\, \sigma_{\rm tot}(E).
\label{ger1}
\end{eqnarray}
This conjecture due to Gerasimov \cite{GER} has often been interpreted as the 
``cancellation between retardation effects
and higher order multipoles'' \cite{FUI},\cite{YAZ}-\cite{ARD}, since the l.h.s. of (\ref{ger1}) 
involves only the unretarded E1 multipole, while the r.h.s. involves the total cross section. It has, however, 
been shown explicitly in ref.
\cite{ARD} that the assumption (\ref{ger}) on the high energy behaviour of the resonance amplitude
is not valid due to the contribution of the anomalous magnetic moment term to the
dispersion integral. Therefore it was pointed out that eqs. (\ref{ger}) and (\ref{ger1}) 
are actually not valid.

\end{document}